\documentclass[11pt, a4paper]{article}
\usepackage{amsmath, amssymb, amsthm}
\usepackage{graphicx}
\graphicspath{{figures/}}
\usepackage{booktabs}
\usepackage{tabularx}
\usepackage[margin=1in]{geometry}
\usepackage{hyperref}
\hypersetup{bookmarksdepth=1}
\usepackage{bm}
\usepackage{empheq}
\usepackage{float}
\usepackage[section]{placeins}  
\usepackage[font=small,labelfont=bf]{subcaption}

\makeatletter
\renewcommand\paragraph{\@startsection{paragraph}{4}{\z@}%
  {3.25ex \@plus1ex \@minus.2ex}%
  {-1em}%
  {\normalfont\normalsize\itshape}}
\makeatother

\title{\textbf{Observable Matrix Dynamics of Stocks}}

\author{Igor Halperin\thanks{All
calculations, numerical analysis, and manuscript
preparation were performed by Claude Code with Opus 4.8
working as an AI assistant under author's supervision. I
would like to thank Miquel Noguer i Alonso, Ernest Baver,
Eric Berger, Galin Georgiev, Andrey Itkin, Charles Martin,
Rudy Martin, Yinsen Miao, and Alejandro Rodriguez Dominguez
for very helpful discussions and comments on the manuscript. All remaining errors are my own. All Python
code, analysis scripts, figures, and animation videos
supporting this paper are available at
\url{https://github.com/ighalp/omd_finance}.}}

\date{\today}

\begin{document}

\emergencystretch=2em

\maketitle

\begin{center}
Email: ighalp@gmail.com
\end{center}

\begin{abstract}
The Observable Matrix Dynamics (OMD) approach monitors the time
development of complex non-linear systems through the trajectory of a
fixed-size distance matrix and its spectrum. We apply it to the
S\&P 500 cross section over three crisis decades, the 2001 dot-com
bust, the 2007--2008 financial crisis, and the 2020 Covid crash, with
three fixed-size observables on a fixed universe. The arccos distance
matrix of the rolling return correlations reads the correlation
geometry: its effective dimension collapses at the 2008 and 2020
crises, while the 2001 bust is a dispersed unwind. Read against
machine-learning distance matrices, its spectrum stays in the
un-relaxed, pre-learning regime with no low-dimensional manifold, so
the market never learns its correlation structure or relaxes to a
stationary geometry. Subtracting the
market factor exposes a coherent sector rotation, whose name-level
attribution identifies which stocks drive each crisis and in what
order. At a short lookback these signals resolve precursors and
forecast the endogenous 2008 crisis, though not the exogenous 2020
shock. The other two observables model the daily return and volatility
rankings as Markov chains on their ranking spaces. The
return chain has persistent, defensive-led bellwethers and
near-reversible dynamics. The volatility chain is far more
persistent, led by the financial sector, and is the only one to carry
a weak, episodic arrow of time, flaring at market stress and matching
volatility clustering and the Zumbach effect. All three matrices show coherent changes during
market crashes.
\end{abstract}

\section{Introduction}
\label{sec:intro}

The eigenvalue spectrum of the equity correlation matrix is
one of the oldest applications of random matrix theory outside
physics. The bulk of the spectrum is well described by the
Marchenko--Pastur law for a noisy sample covariance
\cite{marchenko1967}, while a small number of large eigenvalues
sit above the bulk and carry the market and sector structure
\cite{laloux1999,plerou2002}. A modern textbook account of random
matrix theory for physicists, engineers, and data scientists, which
beyond the general theory devotes an extended treatment to the tools
that analyze financial correlation matrices, is given by Potters and
Bouchaud \cite{potters2020}. During a crisis the cross section
concentrates: correlations rise, the largest eigenvalue grows,
and the effective number of independent factors drops
\cite{onnela2003,kritzman2011}. These facts are usually stated
for a single correlation matrix, or for a slowly varying one,
and read off from its eigenvalues.

Observable Matrix Dynamics (OMD) takes a different vantage
\cite{halperin2026omd}. Its central idea is to map the dynamics of a
complex, high-dimensional system with stochastic and non-linear
evolution onto a single fixed-size object whose own updates carry the
time dimension. The object compactifies the history into a fixed-size
representation, produces a rich set of spectral and geometric diagnostics, and
is universal across tasks, the same construction serving physical and
learning systems alike. Mapping a variable-length input onto a fixed-size internal
representation is a powerful and universal idea. For a vector-valued
internal representation its power is supported by the success of
recurrent neural networks, which fold an unbounded input history into
the fixed-size state carried by their recurrent updates. OMD
proposes a different fixed-size representation by an $N\times
N$ matrix of pairwise distances: distances between particles in
physical systems, or distances between the images (the internal
representations) of $N$ inputs to a neural network in
machine-learning applications. For a
market the particles are the stocks and the distances are the angular
separations of their returns.

In this reading OMD treats the whole trajectory $\{M(t)\}_t$ of the
fixed-size distance matrix as the object of study. Its spectral
reading rests on Inference-BBS (I-BBS)
\cite{halperin2026IBBS}, which infers a latent low-dimensional
sub-manifold from a distance matrix by matching its spectrum to
the Bogomolny--Bohigas--Schmit (BBS) theory of random distance
matrices on Euclidean and spherical spaces
\cite{bogomolny2003,bogomolny2007}, itself part of the Euclidean
random matrix program \cite{mezard1999,goetschy2013}. OMD carries
that inference along a trajectory and asks for quantities intrinsic
to the trajectory rather than to any single snapshot, in the
spirit of the frustrated-distance-matrix (FDM) relaxation dynamics
\cite{halperin2026FDM}. The informative signals are spectral
reorganizations that a scalar loss does not see, together with
trajectory-level observables that link successive snapshots.

This paper applies the OMD reading to financial markets through
three fixed-size matrix observables that give complementary views of
the same daily history. The first is the arccos distance matrix
$M_{ij}(t)=\arccos C_{ij}(t)$ (see Eq.~\eqref{eq:M_def} below) built from a
rolling return correlation window, the geodesic that places the names
on a hyper-sphere $S^{d-1}$. It reads the cross section through its correlations and is
dominated by the market factor, the first principal component of the
return correlation matrix. We read its spectrum for the effective
dimension and the market-factor share, subtract the market factor to
expose the sub-market sector geometry, and track its eigenbasis
rotation with a projector drift and a matrix commutator.

The other two observables are rank-ordered transition matrices.
Ranking the names each day by their running return, or by their
running volatility, turns the relative ordering into a Markov chain
whose transition matrix is a fixed-size observable, one for
performance and one for risk.\footnote{Treating the ordering as a
Markov chain is a first approximation. The short memory and the
absence of covariates can be relaxed by adding sector and
firm-specific factors and higher-order memory as exogenous drivers of
the transition matrix, an extension discussed in
Section~\ref{sec:summary}.} Ranking is market-neutral by
construction, since a common move leaves the order unchanged, so
these chains isolate the relative dynamics that the market factor
hides in $M(t)$. We read their spectra for persistence and mixing,
their entropy production for time-reversibility, and their transfer
entropy for the directed network of who moves whom. We track all
three matrices across three crisis-containing decades and ask
whether the trajectory view adds to the familiar eigenvalue story,
and whether the observables agree.

The distance matrix adds three things beyond a single correlation
snapshot. First, a short lookback sharpens the onset and exposes
precursors that the standard two-year window averages away. Second,
removing the market factor reveals that the sub-market sector geometry
reorganizes on its own schedule, which for the Covid period runs
well past the initial crash. Third, the trajectory-level diagnostics
separate a rotation of the eigenbasis from a mere growth of an
eigenvalue, and show that the market direction is nearly fixed
while the sector direction rotates coherently, with a composition
that identifies each crisis by industry.

A comparison with the machine-learning setting the observable comes
from sharpens the interpretation. In the OMD experiments the
distance-matrix spectrum acquires a low-dimensional manifold structure
only as the network learns. The market spectrum keeps the smooth,
structureless form of an un-relaxed, high-dimensional representation
cloud throughout, with no sign that its correlation structure settles
onto such a manifold. In this reading the market never learns or
relaxes to a stationary geometry, and is better described as a
non-equilibrium object.

The ranking chain adds a market-neutral view of the same history.
Its transition matrix has persistent leaders and laggards beyond
what noise produces, and its migrations are coherent within sectors.
Two further readings follow from the same discrete chain. Its
entropy production, read from the forward and backward transition
matrices, measures the time-irreversibility of the relative dynamics
and finds the return ranking reversible throughout and a weak,
episodic arrow of time in the volatility ranking. The volatility arrow
is near zero in calm markets and flares at market-stress episodes,
strongest at the 2002 dot-com bottom and the 2008 crisis. It is
selective across the stress calendar we examine below, firing on
sustained directional episodes but not on brief single-session shocks,
so it tracks sustained reshuffling rather than volatility itself. Its transfer
entropy gives a directed sector network of who moves whom, and finds
the defensive sectors leading the market through a crisis. Ranking by volatility
instead of return turns the same machine into a risk view, and the
two transition matrices, one for performance and one for risk, are a
matrix-form risk-return analysis in which the risk chain is far more
persistent, the spectral signature of volatility clustering. The
matrix views are
consistent where they overlap and complementary where they do not.
Both see the sector structure tighten toward 2020, and the utilities
block that is the persistent residual cluster of the distance matrix
is the same one that leads the information flow of the ranking chain
in every crisis. A crash appears in the distance matrix as rising
correlation and collapsing dimension and in the ranking as a burst
of migration and irreversibility, the co-movement and the dispersion
being two faces of the same move.

The third period, the Covid decade of 2015--2024, carries a run of
well-documented market-stress events, and we use them to test how the
spectral and matrix signals respond to shocks of different kinds. The
events are the 2021 Archegos liquidation, the 2021 Omicron scare, the
2022 Russian invasion of Ukraine, the 2023 Silicon Valley Bank
failure, the 2024 yen-carry unwind, and the 2024 hawkish-Fed
repricing. The response is selective, and which signal moves
identifies the kind of shock: the arrow of time fires on the
directional deleveraging of Archegos, the spectral concentration on
the correlation shift around the bank failure, while the brief
symmetric shocks leave little trace in either.

The ranking chains are built from discrete states and read as Markov
chains for a reason. Sorting the names into a fixed set of rank
classes and treating the one-step class transitions as a Markov chain
is a deliberate first approximation. Memory beyond one step and
dependence on covariates are both real and can be added as extensions
of this basic setting, yet the memoryless discrete chain already
carries the effects we read here. Its advantage is estimation. The
information-theoretic quantities we want, the entropy, the transfer
entropy, and the entropy production, are hard to estimate for
continuous, high-dimensional data. There they rely on nearest-neighbor
or kernel density estimators whose bias and variance are delicate
\cite{paninski2003}, and dedicated software exists to manage precisely
this difficulty \cite{buth2025infomeasure,wollstadt2019idtxl}. On a
discrete chain with a handful of classes the same quantities reduce to
exact plug-in sums over an estimated transition matrix, cheap to
compute and with a finite-sample bias that is easy to control against
a reversible null. The numbers this produces are well defined and
carry an intuitive reading, as the analysis below shows.

\paragraph{Related work.}
The noise-dressing of financial correlation matrices and the
random matrix approach to cross correlations established the
bulk-plus-outliers picture \cite{laloux1999,plerou2002}, and
hierarchical and network methods track how the correlation
structure evolves in time \cite{onnela2003}. The absorption
ratio, the fraction of variance in the leading principal
components, is a systemic-risk indicator close to the
participation ratio used here \cite{kritzman2011}. The
distance-matrix and trajectory-level tools we use are taken
from the frustrated-distance-matrix study \cite{halperin2026FDM}
and the OMD framework \cite{halperin2026omd}, which in turn
build on \cite{bogomolny2003,mezard1999,goetschy2013}. The
contribution here is empirical: to run those tools on real
equity data across three crises and read what they add.

\section{Data and the time-dependent distance matrix}
\label{sec:data}

The data are daily CRSP records for S\&P 500 constituents over
1996--2024. For
a chosen ten-year period we take the set of names that are
index members for the entire period, including the two-year
pre-roll needed to fill the first lookback, so that the cross
section does not drift inside the period. We build the wide
panel of daily total returns on that fixed universe and slide a
lookback window of fixed length $L$ in daily steps. At each
measurement date $t$ the correlation matrix $C(t)$ is the
Pearson correlation over the last $L$ observations. The Pearson
correlation is the cosine-similarity Gram of the demeaned,
unit-normalized return vectors, $C_{ij} = \cos\theta_{ij}$ with
unit diagonal, so the distance matrix is the arccos of that Gram
taken element-wise,
\begin{equation}
\label{eq:M_def}
M_{ij}(t) = \arccos G_{ij}(t), \qquad
G_{ij}(t) = \frac{C_{ij}(t)}{\sqrt{C_{ii}(t)\,C_{jj}(t)}},
\end{equation}
the geodesic angle between the return directions on the sphere.
For a raw correlation matrix $G = C$, so $M = \arccos C$; the
renormalization to a unit-diagonal cosine Gram matters only when
the input is not already a correlation, as for the deflated
residual of Section~\ref{subsec:mktrm}. Here $M_{ij}\in[0,\pi]$
with a vanishing diagonal. Fixing the
number of observations $L$ keeps the aspect ratio $q=N/L$
constant in time, so the sampling noise in the spectrum is the
same at every date and spectra are comparable across the
period.

We study three periods, each centered on a crisis: 1996--2005
for the 2001 dot-com bust, 2003--2012 for the 2007--2008
financial crisis, and 2015--2024 for the 2020 Covid crash. The
constant-membership universe holds $N=244$, $265$, and $309$
names respectively at the two-year lookback. The
default lookback is $L=504$ trading days, about two years.
When we shorten the lookback we cap the universe to the largest
names by average market capitalization so that $N<L$ with
margin, keeping $q\approx 0.75$ and the correlation estimate
out of the singular $N>L$ regime: $N=94$ at $L=126$ (about six
months) and $N=192$ at $L=256$ (about one year).

\section{M-matrices in finance versus machine learning and physics}
\label{sec:observables}

The distance matrix $M(t)$ that this paper tracks is borrowed from two
recent settings, machine learning in the OMD and I-BBS papers
\cite{halperin2026omd,halperin2026IBBS}, and statistical physics in
the frustrated-distance-matrix model \cite{halperin2026FDM}. In both
the object was inspired by the Bogomolny--Bohigas--Schmit theory of
random distance matrices \cite{bogomolny2003,bogomolny2007}. BBS
itself is not concerned with dynamics. It describes the spectrum of
the distance matrix of a single static snapshot of points distributed
uniformly on the surface of a hyper-sphere. The OMD and FDM
constructions carry that static object along a trajectory, and in both
the underlying system relaxes to an equilibrium configuration, a
trained representation or a settled particle arrangement. The BBS
spectral characteristics, above all the delocalisation slope
$\beta(t)$, then serve as a running diagnostic of that relaxation. A
jump in $\beta(t)$ marks a structural change, and a stage with
$\beta>1$ signals the emergence of an effective low-dimensional
manifold, as at the grokking transition of the OMD paper where the
representation settles onto its learned sub-manifold.

The equity cross section, as the rest of this section shows, does not
relax in this way, and $\beta$ becomes a secondary diagnostic rather
than a signature of learning. We first set out the spectral
observables, then collect the caveats that keep the strict BBS
structure from applying here, and finally read the finding against the
machine-learning and physical cases.

We read two companion spectra at each date, following the
I-BBS treatment of a random distance matrix. The correlation
matrix $C(t)$ is the cosine-similarity Gram of the return
directions on the sphere, and its ordered eigenvalues
$\lambda_1\ge\lambda_2\ge\cdots$ give the standard factor
observables: the market-factor share $\lambda_1/\sum_k\lambda_k$ and
the participation ratio
\begin{equation}
\label{eq:pr}
\mathrm{PR}(t) = \frac{\bigl(\sum_k \lambda_k\bigr)^2}
{\sum_k \lambda_k^2},
\end{equation}
which counts the effective number of factors.

The distance matrix $M(t)$ adds a second reading, following the same
I-BBS decomposition into a Perron mode and a power-law branch
\cite{bogomolny2003,halperin2026IBBS}. The Perron eigenvalue
$\Lambda_1(t)$, the single large positive eigenvalue of the
angular-distance matrix, tracks the overall scale, the mean angular
separation of the cross section. The largest-magnitude non-Perron
eigenvalues on the window $K\in[2,\sqrt N]$ form a delocalised branch
that decays as $|\lambda_K|\sim K^{-\beta}$, which for a uniform sample
on the hyper-sphere $S^{d-1}$ would give $\beta=d/(d-1)$ and an
effective embedding dimension $d_\beta=\beta/(\beta-1)$. We fit $\beta$
on that window directly from the eigenvalues of $M(t)$.

\begin{figure}[htbp]
\centering
\includegraphics[width=\textwidth]{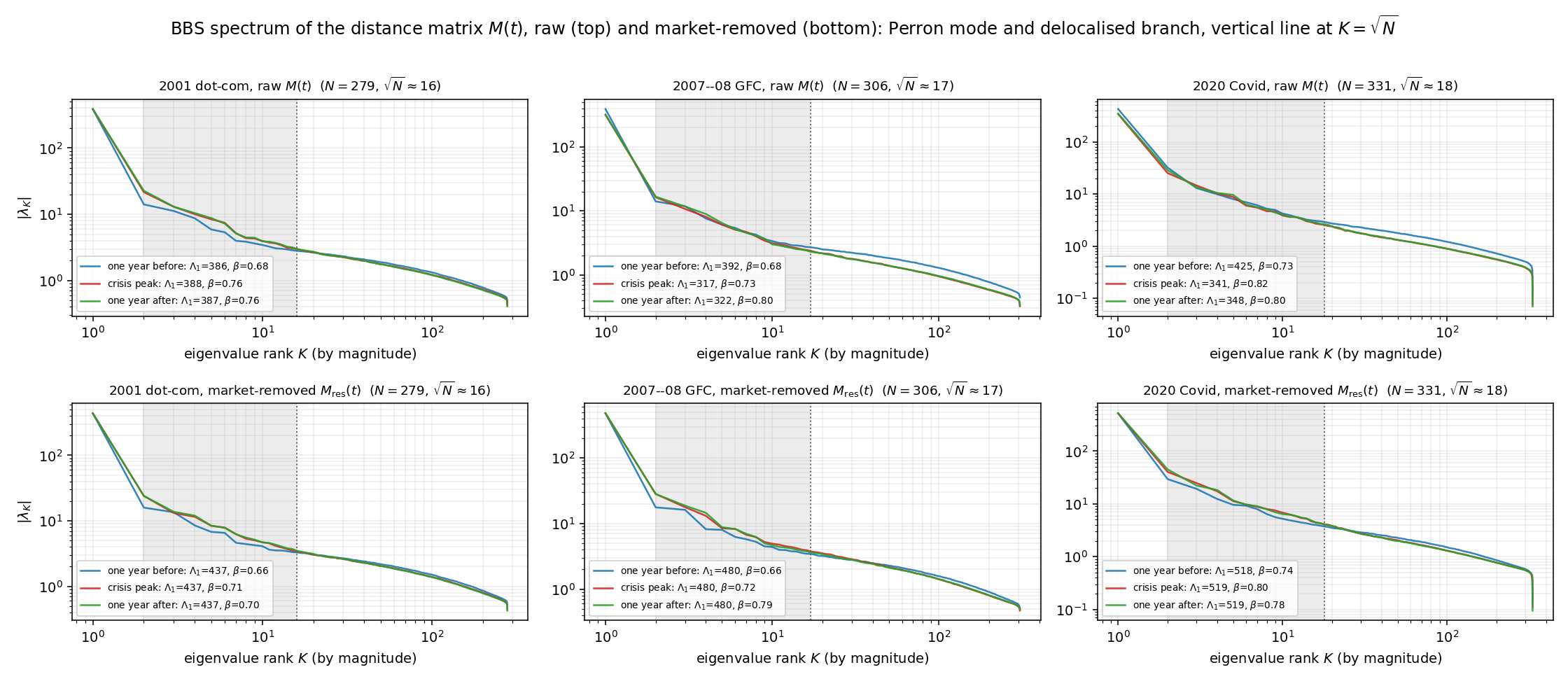}
\caption{BBS spectrum of the distance matrix $M(t)$, two-year
lookback, raw (top) and market-removed (bottom), one column per
period. Each panel shows the eigenvalue magnitudes $|\lambda_K|$
ranked on a log-log scale at three phases: one year before onset
(blue), the crisis peak (red), and one year after (green). The Perron
eigenvalue $\Lambda_1$ is the isolated point at rank one, and the
delocalised branch $|\lambda_K|\sim K^{-\beta}$ decays over the shaded
window $K\in[2,\sqrt N]$ where $\beta$ is fit. The dotted vertical
line marks $K=\sqrt N$, where the BBS shoulder between the delocalised
and the localised branch would sit. The legend gives $\Lambda_1$ and
$\beta$ per curve.}
\label{fig:bbsspectrum}
\end{figure}

Figure~\ref{fig:bbsspectrum} shows the spectrum itself, the
eigenvalue magnitudes of $M(t)$ ranked on a log-log scale at three
phases of each crisis, one year before onset, the crisis peak, and
one year after, for the raw distance matrix (top) and the
market-removed residual (bottom). The single Perron eigenvalue
$\Lambda_1$ sits well above the rest at rank one, and the delocalised
branch decays as the power law $|\lambda_K|\sim K^{-\beta}$ over the
shaded window $K\in[2,\sqrt N]$. The crisis reads from the scale. For
the 2008 and 2020 crashes the raw Perron contracts at the peak, from
$392$ to $317$ and from $425$ to $341$, as the rising correlations
shrink the angular separations, while the 2001 dispersion leaves it
nearly unchanged. Removing the market factor stabilizes the scale, the
residual Perron holding near $480$ for 2008 and $519$ for 2020 across
all three phases, which locates the contraction in the market factor
itself. The delocalised exponent is $\beta\approx0.7$ throughout and
steepens slightly at the peak in both readings.

Read literally, these spectra are a negative result for the strict BBS
theory, and they differ in the same way from the I-BBS and OMD spectra
obtained in machine-learning experiments and the FDM spectra obtained
in physical systems. In all of those the delocalised branch resolves
into flat angular-momentum quasi-multiplets, and a shoulder separates
it from the localised bulk near $K\approx\sqrt N$.
Our spectra show neither: the decay is smooth and monotonic, with no
multiplet plateaus and no shoulder at the dotted $K=\sqrt N$ line, and
removing the market factor does not recover the structure. Three
effects plausibly wash it out. The lookback is finite with $T$ of
order $N$, so the correlation carries a Marchenko--Pastur noise bulk
\cite{marchenko1967} that smears the fine structure a noiseless
manifold sample would show. The universe is small,
$\sqrt N\approx16$ to $18$, so the delocalised window holds too few
modes to resolve a shoulder. And the cross section is a
few-factor-plus-noise structure rather than a uniform sample from a
low-dimensional hyper-sphere, without the angular-momentum symmetry
that produces multiplets. We therefore read only the coarse features,
the Perron scale and the delocalised slope $\beta$, and treat the
absence of the fine BBS structure as an open point, to be probed with
longer lookbacks, larger universes, or noise-cleaned correlation
estimators.

For the equity cross section the delocalised exponent stays below
one throughout, near $\beta\approx0.7$, and steepens toward one as a
crisis concentrates the cross section. The naive inversion
$d_\beta=\beta/(\beta-1)$ then turns negative, which is not a physical
embedding dimension. The delocalised reading is asymptotic in the
sample size, valid in the BBS regime $N\gg d^2$, and at finite $N$ the
measured exponent falls below its large-$N$ asymptote $d/(d-1)$ by an
I-BBS finite-$N$ correction $\Delta\beta(N,d)$ that grows as $N$
shrinks toward $d^2$ \cite{halperin2026IBBS}. At our fixed $N$ of a
few hundred names, $\beta<1$ is therefore consistent either with a
finite latent dimension once that correction is restored, or with an
embedding whose points depart from a uniform sample on the latent
hyper-sphere. The market factor is a natural source of such
non-uniformity, but re-measuring $\beta$ on the market-removed
distance matrix leaves it near $0.7$, so the shortfall below one
reflects the finite-$N$ limitation at our fixed $N$ rather than a
market-factor artifact. We therefore report the participation ratio of
$C(t)$ as the primary effective-factor count, with the Perron
eigenvalue and the delocalised $\beta$ of $M(t)$ as a secondary
reading alongside it. We take $\beta$ as an empirical power-law slope
rather than as evidence that the underlying BBS geometry has formed.

\begin{figure}[htbp]
\centering
\includegraphics[width=\textwidth]{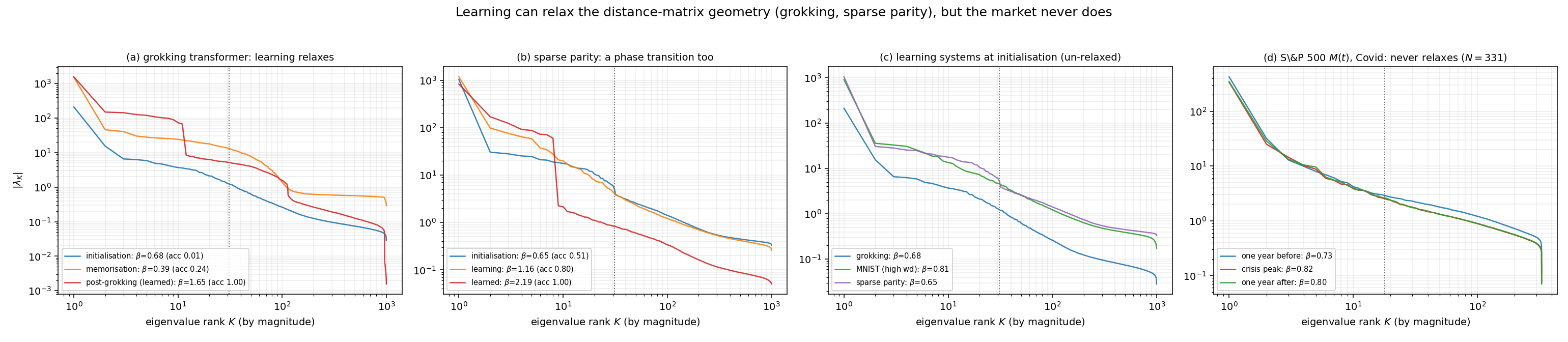}
\caption{A learning system can relax its distance-matrix geometry, and
the market does not. (a) The grokking transformer of
\cite{halperin2026omd}: the train distance-matrix spectrum at
initialisation, memorisation, and after grokking, the exponent rising
from $\beta\approx0.68$ through $0.39$ to $\beta\approx1.65$ as a
shoulder near $K=\sqrt N$ (dotted) emerges. (b) Sparse parity undergoes
the same phase transition more sharply, from $\beta=0.65$ to $2.19$.
(c) At initialisation grokking, MNIST, and sparse parity share the
un-relaxed regime, $\beta$ between $0.65$ and $0.81$. Not all then
relax through a transition, MNIST doing so only diffusively
\cite{halperin2026omd}. (d) The S\&P 500 through the Covid crisis, one
year before, at the peak, and after, its exponent near $0.7$ to $0.8$
with no shoulder, never relaxing.}
\label{fig:learnvsmarket}
\end{figure}

Read through the OMD lens, this same negative finding becomes
constructive, and Figure~\ref{fig:learnvsmarket} makes it concrete. The
un-relaxed, high-dimensional representation cloud of a learning system
has a shared spectral signature at initialisation, before any
training: a smooth power-law decay with $\beta$ between $0.65$ and
$0.81$, no quasi-multiplets, and no shoulder at $K=\sqrt N$. Grokking,
MNIST, and sparse parity all begin with their untrained exponent
$\beta$ in this narrow range (panel c), so the value is a property of
the un-relaxed cloud rather than of any one task, while experiments
with intrinsically low-dimensional inputs start already structured. Learning can change it, but not for every system.
Grokking and sparse parity pass through a geometric phase transition,
the exponent rising to $\beta\approx1.65$ and $2.19$ with a shoulder at
$K=\sqrt N$ (panels a and b), while MNIST relaxes only diffusively,
without one \cite{halperin2026omd}. The market carries the same
un-relaxed signature and never departs from it, its exponent staying
near $0.7$ across the Covid crisis (panel d), with
the crisis trajectories of this paper the direct evidence that the
geometry does not settle. The natural reading is that no learning of
the correlation structure into a low-dimensional manifold takes place
in the market. It never relaxes to an equilibrium $M$-matrix. Its
correlation and distance matrices keep evolving with no stationary
limit, which points to a non-equilibrium description, plausibly a
non-equilibrium steady state (NESS) through periods of steady market
conditions rather than a relaxed equilibrium.

These observations set the plan for the rest of the paper. Because the
market never relaxes onto a low-dimensional manifold, $\beta$ here
plays a smaller role than in the OMD and FDM settings, where it is the
main signature of learning and of relaxation to a steady state. We
keep it as one diagnostic among several, its changes flagging shifts
in the correlation structure alongside the Perron scale and the
effective factor count, rather than as a primary readout. The dynamic
characteristics that the moving $M$-matrix does carry are the subject
of the sections that follow: the spectral collapse and the coherent
eigenbasis rotation of the distance matrix, and the persistence,
entropy production, and transfer entropy of the rank-ordered
transition matrices that read the relative dynamics the market factor
hides.

\section{Results}
\label{sec:results}

Table~\ref{tab:summary} collects the spectral-collapse jumps at each
onset, across three lookback windows, before the subsections develop
them. The 2008 and 2020 crises share a signature, a collapse in the
effective factor count and a jump in the market-factor share, while the
2001 dot-com period runs the other way. The size of the jump depends
on the lookback. A short window resolves a sharper collapse, the Covid
market share rising to $0.66$ over a $126$-day window against $0.48$
over two years, while the 2001 dispersion signature, a rising factor
count and a falling market share, emerges only at the longer windows
and reverses at $126$ days. The protagonist is the sector cluster that
dominates the market-removed eigenvector's rotation through the crisis,
and the last
column records whether the fragility signals forecast the crisis and
the per-period forecast AUC at the 63-day horizon. The table covers
the three crises. The Covid decade additionally offers a calendar of
notable market-stress events, the 2021 Archegos liquidation, the 2021
Omicron scare, the 2022 Russian invasion of Ukraine, the 2023 Silicon
Valley Bank failure, the 2024 yen-carry unwind, and the 2024
hawkish-Fed repricing, against which we test the sensitivity of the
spectral and ranking-chain signals in the analysis that follows.

\begin{table}[htbp]
\centering
\footnotesize
\setlength{\tabcolsep}{4.5pt}
\caption{Summary across the three crises and three lookback windows.
Effective factors is the participation ratio of $C(t)$, and market
share is $\lambda_1/\sum\lambda$, each averaged over the calm year
before onset and the first two months of the crash. The $126$- and
$256$-day windows use a market-cap-capped universe so that $T>N$; the
two-year ($504$-day) window uses the full constant-membership
universe. The jumps sharpen at shorter lookback for the 2008 and 2020
collapses, while the 2001 dispersion signature, a rising factor count
and a falling market share, emerges only at the longer windows. The
last column gives the protagonist sector, the onset shape at the name
level, and the per-period forecast AUC at the 63-day horizon.}
\label{tab:summary}
\begin{tabular}{l c c c c l}
\toprule
Crisis & Lookback & $N$ & Eff.\ factors & Market share
 & Protagonist, \\
 & (days) & & (calm $\to$ crash) & (calm $\to$ crash) & onset forecast \\
\midrule
2001 dot-com & 126 & $94$  & $16.5 \to 14.9$ & $0.20 \to 0.20$
 & Technology, \\
 & 256 & $192$ & $19.8 \to 23.6$ & $0.20 \to 0.15$ & staggered, \\
 & 504 & $244$ & $22.9 \to 25.3$ & $0.19 \to 0.17$ & no (0.46) \\
\midrule
2007--08 GFC & 126 & $94$  & $5.1 \to 3.8$ & $0.43 \to 0.51$
 & Energy, \\
 & 256 & $192$ & $6.1 \to 4.5$ & $0.40 \to 0.47$ & gradual, \\
 & 504 & $265$ & $8.3 \to 5.3$ & $0.34 \to 0.43$ & yes (0.72) \\
\midrule
2020 Covid & 126 & $94$  & $6.6 \to 2.4$ & $0.37 \to 0.66$
 & REITs, utilities, \\
 & 256 & $192$ & $7.3 \to 3.3$ & $0.36 \to 0.57$ & synchronized, \\
 & 504 & $309$ & $9.0 \to 4.4$ & $0.32 \to 0.48$ & no (0.49) \\
\bottomrule
\end{tabular}
\end{table}

\subsection{Spectral collapse onto the market factor}
\label{subsec:collapse}

At the onset of the 2008 and 2020 crises the correlation
spectrum collapses onto the market factor. For the Covid period,
the mean pairwise correlation rises from $0.29$ in the 2019
baseline to $0.47$ in the March--April 2020 window, the
market-factor share rises from $0.31$ to $0.48$, and the
participation ratio falls from about nine effective factors to
about four. The delocalised exponent of $M(t)$ steepens from
$\beta\approx 0.73$ to $\beta\approx 0.79$ as the spectrum
concentrates. The Perron eigenvalue of $M(t)$, the scale mode,
contracts from about $394$ to $335$ as the rising correlations
shrink the angular separations. Because the lookback is two years,
the elevated state persists for about two years and then steps
back down as the crash leaves the window.

\begin{figure}[htbp]
\centering
\includegraphics[width=0.72\textwidth]{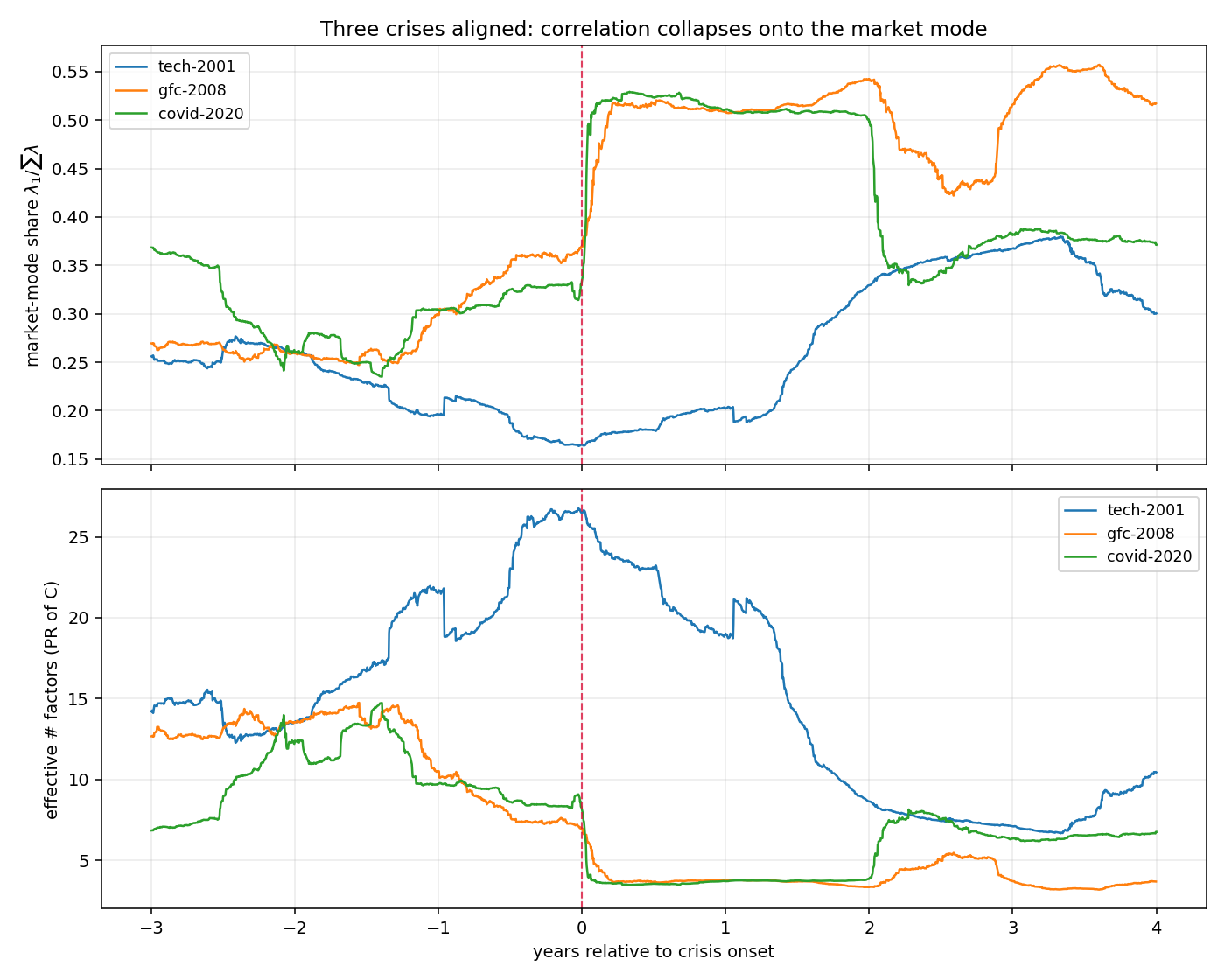}
\caption{Three crises aligned on time-to-onset (years). Top:
market-factor share $\lambda_1/\sum\lambda$ of $C(t)$. Bottom:
effective number of factors, the participation ratio
Eq.~\eqref{eq:pr}. The 2008 and 2020 crises collapse the
factor count and raise the market share at onset; the 2001
dot-com period does the reverse, a dispersed decorrelated
unwind. Two-year lookback.}
\label{fig:aligned}
\end{figure}

\begin{figure}[htbp]
\centering
\includegraphics[width=0.66\textwidth]{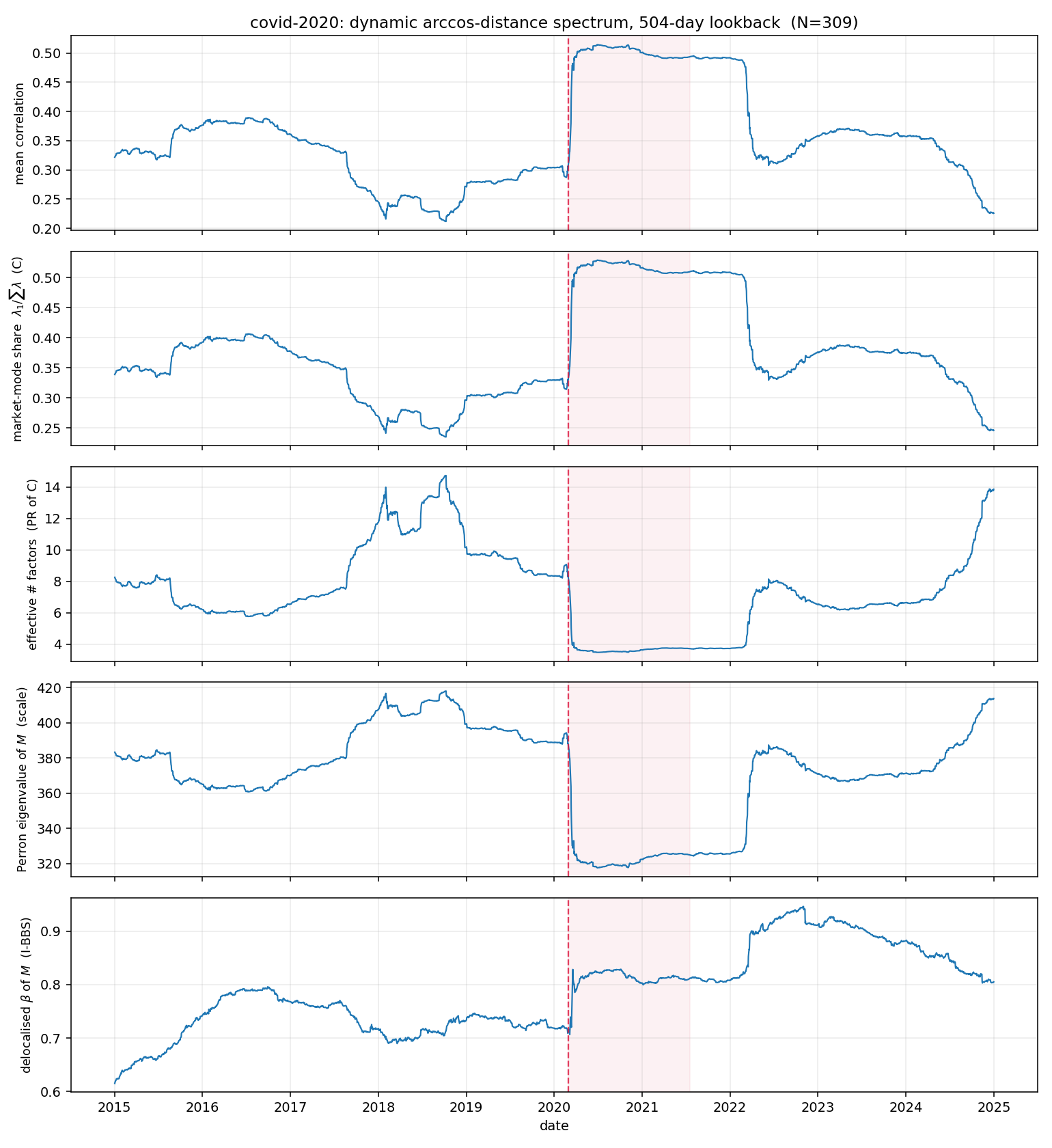}
\caption{Covid period, two-year lookback, $N=309$. From top:
mean correlation, market-factor share and effective factor count of
$C(t)$, then the Perron eigenvalue and the delocalised $\beta$ of
the distance matrix $M(t)$. The dashed line marks the onset and
the shaded band the two-year lookback window that follows it.}
\label{fig:covid}
\end{figure}

Figure~\ref{fig:aligned} aligns the three crises on a common
time-to-onset axis. The 2008 and 2020 crises share the same
signature: a sharp drop in the effective factor count and a jump
in the market-factor share at $t=0$. The 2001 dot-com period is
the opposite. There the participation ratio peaks and the
market-factor share dips near the onset. The bust was a dispersed,
decorrelated unwind in which technology names pulled apart from
the rest while the market as a whole held together. The observable
therefore separates a correlated crash from a dispersion event,
a distinction that a single rising average correlation would
miss. Figure~\ref{fig:covid} shows the full trajectory for the
Covid period, where each observable moves sharply at onset,
plateaus for the lookback length, and reverts, with the 2022
bear market visible as a secondary feature.

\subsection{Lookback dependence and onset localization}
\label{subsec:lookback}

\begin{figure}[htbp]
\centering
\includegraphics[width=0.82\textwidth]{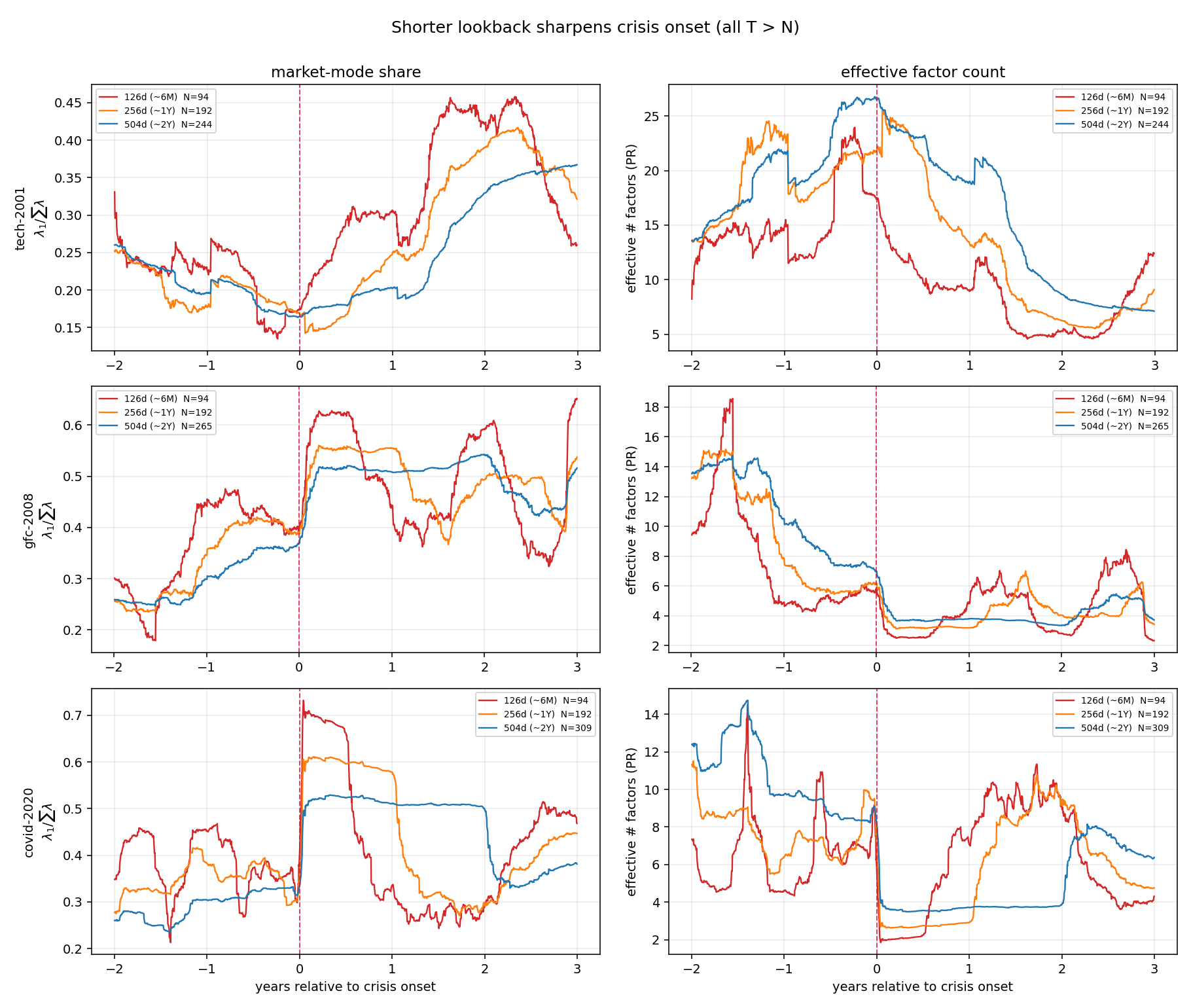}
\caption{Lookback comparison at 126, 256, and 504 days, all in
the $L>N$ regime (the short windows use a market-cap-capped
universe). Left: market-factor share. Right: effective factor
count. Rows are the three crises, aligned on time-to-onset. The
short window sharpens the onset and resolves the mid-2007
tremor that the two-year window smooths out.}
\label{fig:lookback}
\end{figure}

The two-year window smears the crisis over its own length. A
shorter lookback localizes the onset, at the cost of a noisier
correlation estimate, and to keep the estimate out of the
$N>L$ regime we cap the universe as described in
Section~\ref{sec:data}. Figure~\ref{fig:lookback} overlays the
126, 256, and 504-day lookbacks, all with $N<L$. At 126 days
the Covid market-factor share spikes to about $0.73$, against
$0.52$ at 504 days, and the effective factor count drops to
about two, against about four. The short window also decays
within about a year rather than sitting on the mechanical
two-year plateau, tracking the actual recovery more faithfully.
The short lookback resolves structure that the long window
averages away. In the 2008 period the 126-day participation
ratio shows a sharp dip in mid-2007, about a year before the
Lehman failure, that is invisible at two years. This is the
subprime and quant-liquidation tremor, and its early resolution
is a concrete example of information that the trajectory view
adds at short lookback.

\subsection{Removing the market factor}
\label{subsec:mktrm}

To see the structure hidden under the dominant factor we remove
the market factor by spectral deflation of the correlation matrix.
At each window we diagonalize $C(t)$, take the leading eigenpair
$(\lambda_1, V_1)$, the market factor, and subtract its rank-one
projection,
\begin{equation}
\label{eq:deflate}
C_{\mathrm{res}}(t) = C(t) - \lambda_1 V_1 V_1^{T},
\end{equation}
which sends the market direction to a zero eigenvalue and leaves
the residual correlation with leading eigenvalue $\lambda_2$, the
largest sector factor. The market-removed distance matrix is
formed from the residual by the same construction as
Eq.~\eqref{eq:M_def}, renormalizing $C_{\mathrm{res}}$ to a
unit-diagonal cosine Gram and taking the arccos element-wise, and
its spectrum is read in the same I-BBS way. We report the Perron
eigenvalue and delocalised $\beta$ of $M_{\mathrm{res}}$, and the
leading-sector share and participation ratio of $C_{\mathrm{res}}$.

\begin{figure}[htbp]
\centering
\includegraphics[width=0.95\textwidth]{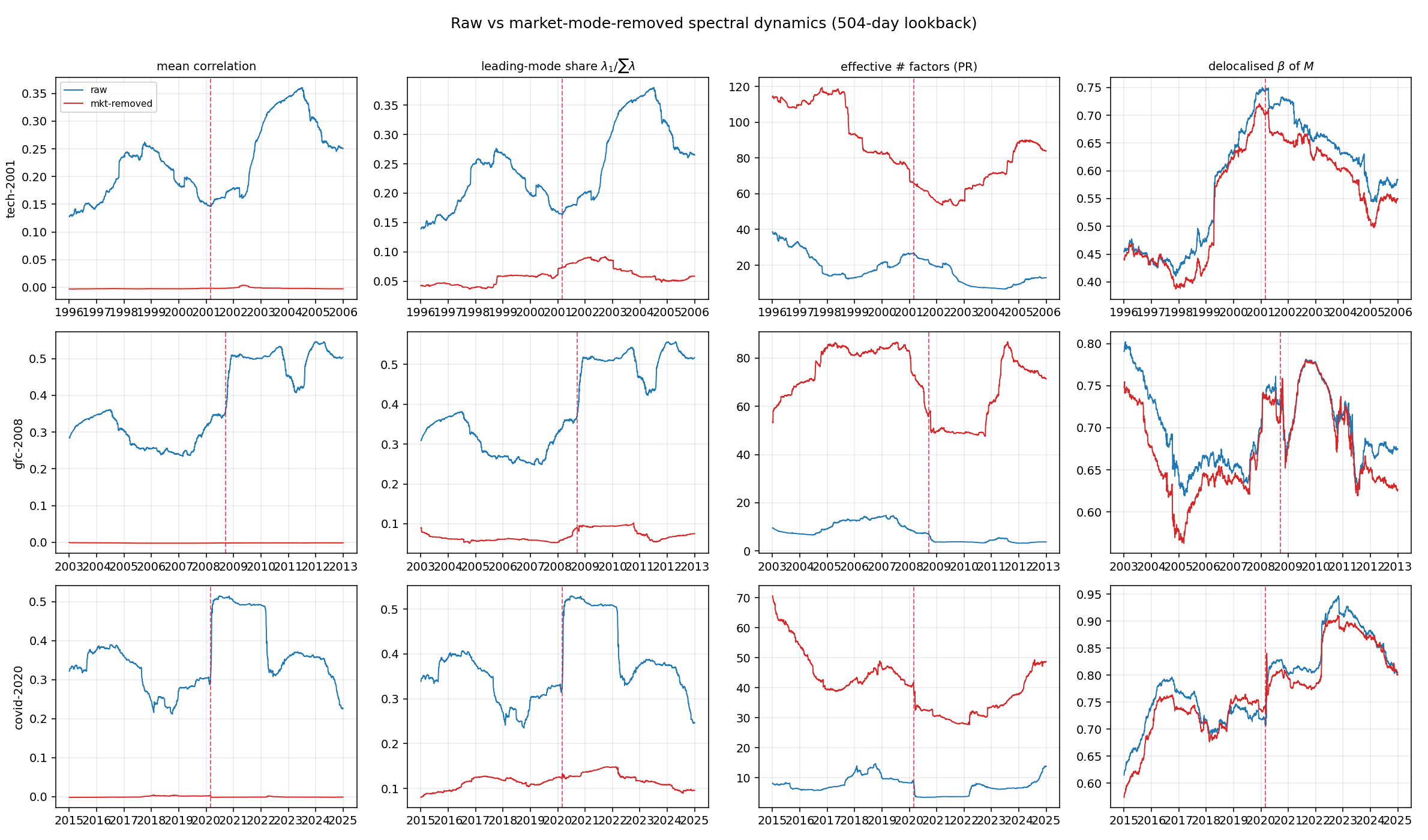}
\caption{Raw versus market-factor-removed spectral dynamics,
two-year lookback. Columns: mean correlation, leading-mode share
and effective factor count of the correlation matrix, and the
delocalised $\beta$ of the distance matrix. Rows are the three
crises. Spectral deflation zeroes the residual mean correlation
and lifts the effective dimension, yet the onset signature
persists in the residual, with a timing that can decouple from
the market.}
\label{fig:mktrm}
\end{figure}

Removal does what it should. The residual mean correlation sits
at zero, the leading-sector share falls to between $0.06$ and
$0.12$, and the effective factor count rises severalfold, from
roughly seven to eighteen in the raw matrix to forty to
eighty-five in the residual. The market factor dominates the raw
geometry. The crisis signature nonetheless
survives the removal. At each onset the residual leading-sector
share still rises, the residual participation ratio still
falls, and the residual $\beta$ still steepens.
Figure~\ref{fig:mktrm} shows the comparison. A crisis therefore
reaches beyond a market-beta event: the sub-market sector
co-movement also tightens, though far more mildly than the
market factor. The timing can decouple. For the Covid period the
residual leading-sector mode reaches its strongest
concentration in 2021--2022, well after the March 2020 crash,
tracking the post-crash sector rotation that the market factor
does not see.

\subsection{Eigenvector rotation through the crises}
\label{subsec:rotation}

The eigenvalue observables above do not distinguish a rotation
of the eigenbasis from a growth of an eigenvalue at fixed
direction. To separate the two we use two trajectory-level
diagnostics, the top-$K$ projector
drift\footnote{We thank Alejandro Rodriguez Dominguez for
suggesting these two trajectory-level metrics.}
\begin{equation}
\label{eq:Dk}
D_K(t_{\mathrm{ref}}, t) =
\bigl\| P_K(t) - P_K(t_{\mathrm{ref}}) \bigr\|_F,
\qquad P_K(t) = V_K(t) V_K(t)^{T},
\end{equation}
where $V_K(t)$ holds the top-$K$ eigenvectors of $C(t)$, and
the Frobenius norm of the matrix commutator
\begin{equation}
\label{eq:comm}
\mathcal{C}(t_{\mathrm{ref}}, t) =
\bigl\| [\, C(t_{\mathrm{ref}}), C(t)\, ] \bigr\|_F .
\end{equation}
$D_K$ measures whether the top-$K$ eigenspace rotates between
the two times, and $\mathcal{C}$ measures whether the two
matrices share an eigenbasis, vanishing when they commute and
growing with the amount of basis rotation. In the
frustrated-distance-matrix study the informative eigenspace was
the bottom-$K$ block of the distance matrix
\cite{halperin2026FDM}; here the market and sector structure
sits in the top-$K$ block of the correlation matrix, so we
track the top-$K$ rotation. We anchor at a pre-crisis reference,
one year before onset, and follow the diagnostics as the window
slides through the crisis. The onset itself is a fixed, hand-chosen
date for each period, the dot-com decline in March 2001, the Lehman
collapse in September 2008, and the Covid crash in March 2020. By
construction $D_K$ and $\mathcal{C}$ vanish at the anchor and rise as
the eigenbasis rotates, so the drift minimum at the anchor (the green
line) is a reference point where the measurement is zeroed, not a
year-ahead precursor of the crisis. To judge whether a rotation is
coherent or random, we compare the diagnostics against a
random-subspace null, the level each would reach if the top-$K$
eigenspace at time $t$ were oriented at random relative to the anchor.
For the projector drift, two independent random $K$-frames $V_1, V_2$
in $\mathbb{R}^N$ give
$\|P_1-P_2\|_F^2 = 2\,(K - \|V_1^{T} V_2\|_F^2)$, and
$\mathbb{E}\,\|V_1^{T} V_2\|_F^2 = K^2/N$, so the expected drift is
$\sqrt{2K(1-K/N)}$, about $3.1$ for $K=5$ and $N$ of a few hundred.
For the leading-eigenvector overlap, two independent random unit
vectors in $\mathbb{R}^N$ have root-mean-square overlap $1/\sqrt{N}$.
These are the analytic analog of the eigenvalue-randomized null of
\cite{halperin2026FDM}. A drift near $\sqrt{2K(1-K/N)}$ or an overlap
near $1/\sqrt{N}$ marks a fully scrambled eigenspace, and values far
from those levels a coherent, partial rotation.

\begin{figure}[htbp]
\centering
\includegraphics[width=0.98\textwidth]{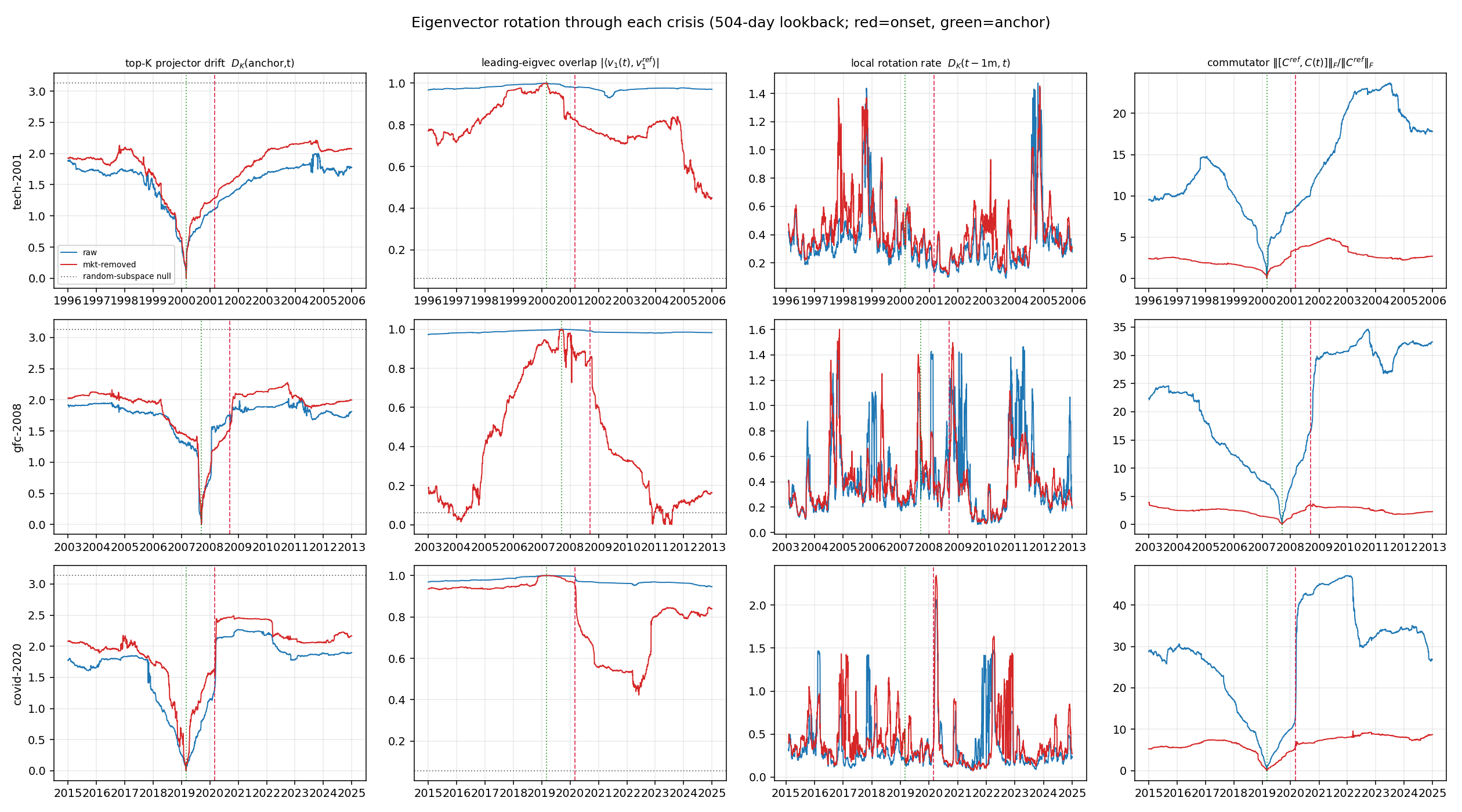}
\caption{Eigenvector rotation through each crisis, two-year
lookback. Blue is the raw correlation matrix, red the
market-removed residual. Columns: top-$K$ projector drift $D_K$ from
the anchor (see Eq.~\eqref{eq:Dk}), leading-eigenvector overlap with
the anchor, the one-month local rotation rate, and the commutator
norm $\mathcal{C}$ (see Eq.~\eqref{eq:comm}). The vertical lines mark
the anchor, one year before onset (green dotted), and the onset
(crimson dashed). The horizontal grey dotted line in the first two
columns is the random-subspace null defined in the text. The raw
market eigenvector holds overlap near one, while the market-removed
sector eigenvector rotates and the commutator steps up at onset.}
\label{fig:rotation}
\end{figure}

\begin{figure}[htbp]
\centering
\includegraphics[width=0.62\textwidth]{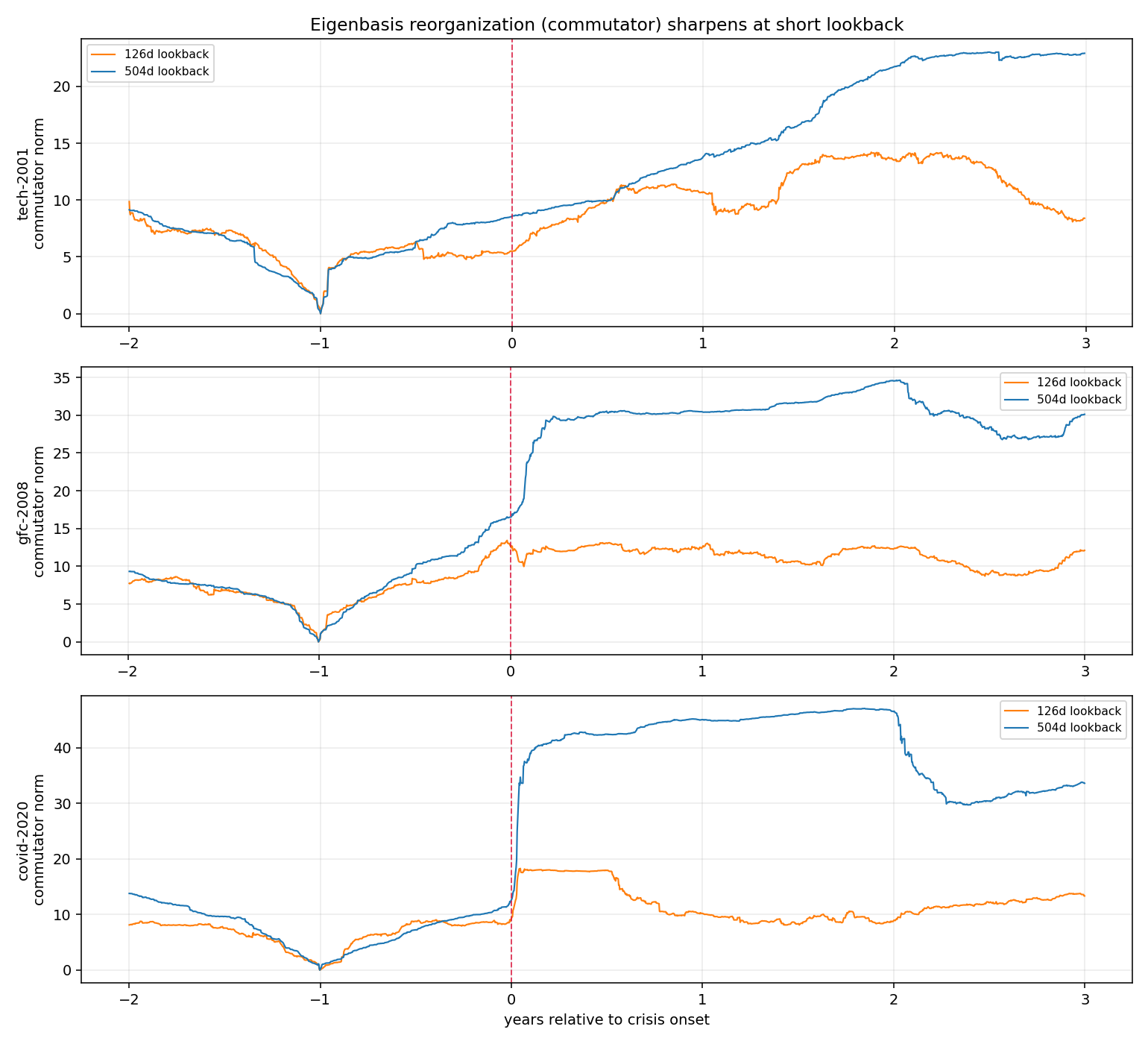}
\caption{Eigenbasis reorganization measured by the commutator
norm $\mathcal{C}(t_{\mathrm{ref}}, t)$, 126 versus 504-day
lookback, aligned on time-to-onset. The short window localizes
the reorganization to the onset and decays within about a year;
the long window shows the mechanical two-year plateau.}
\label{fig:comm}
\end{figure}

Figure~\ref{fig:rotation} reports the diagnostics at the
two-year lookback. The market direction amplifies while its
orientation holds fixed. The raw leading eigenvector keeps an
overlap near one,
between $0.93$ and $0.99$, with its pre-crisis anchor through
every crisis: the market factor grows in eigenvalue but keeps its
direction, which is the nearly uniform positive combination of all
names. The sector geometry is what reorganizes. With the market
removed the leading sector eigenvector rotates strongly, its
anchor overlap falling to about $0.5$ for Covid and below $0.2$
for 2008, and for the Covid period it keeps rotating through the
2020--2022 sector rotation rather than only at the March 2020
crash. The commutator gives a sharp
onset marker complementary to the eigenvalue collapse: the norm
$\mathcal{C}(t_{\mathrm{ref}}, t)$ steps up abruptly at each
onset, from zero at the anchor to about $43$ for Covid and about
$30$ for 2008, as the correlation matrix ceases to commute with
its pre-crisis self. At the 126-day lookback the step localizes
to the onset and decays within about a year, against the
mechanical two-year plateau of the long window
(Figure~\ref{fig:comm}). Throughout, the top-$K$ projector
drifts stay well below the random-subspace null. The rotation is
coherent, the same conclusion that
\cite{halperin2026FDM} reaches for its relaxation trajectory:
the matrix trajectory carries information beyond the
instantaneous eigenvalue sequence.

\subsection{Which sectors the rotating eigenvector loads onto}
\label{subsec:sectors}

The rotating market-removed leading eigenvector is a sector
portfolio, and its composition names the crisis. We classify
each name into one of about ten sectors from its SIC code, a
SIC-derived grouping in which pharmaceuticals and computing are
pulled out of the broad chemicals and manufacturing groups. At
each date we take the energy fraction of the leading residual
eigenvector in each sector, $\sum_{i\in s} v_{1,i}^2$, and
normalize by the sector's share of names to a concentration
ratio, so that a value above one means the eigenvector
over-weights that sector.

\begin{figure}[htbp]
\centering
\includegraphics[width=0.98\textwidth]{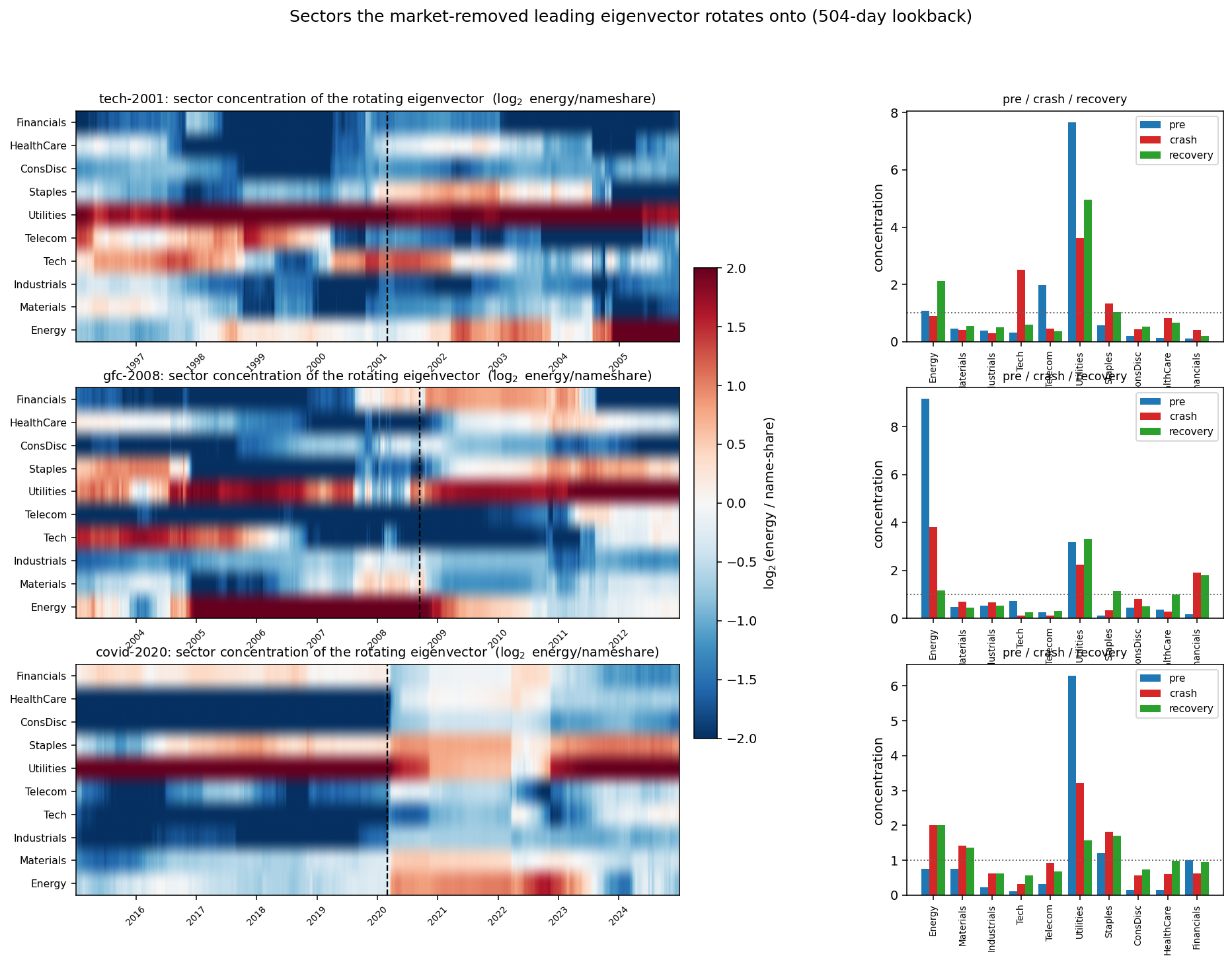}
\caption{Sector composition of the rotating market-removed
leading eigenvector, two-year lookback. Left: heatmap of the
concentration ratio, a sector's fraction of the eigenvector divided
by its share of names, in $\log_2$ units, with onset
dashed. Right: the same ratio at a pre-crisis, crash, and
recovery date. Utilities is persistently the most concentrated
cluster. On top of it, technology marks the 2001 period, the
energy supercycle the 2003--2012 period with financials rising in
2008, and energy the 2020 period. The sectors are SIC-derived, so
the real estate investment trusts that drive the 2020 name-level
rotation (Figure~\ref{fig:names}) sit inside the Financials row here,
whose concentration falls together with the utilities as energy
rises.}
\label{fig:sectors}
\end{figure}

Figure~\ref{fig:sectors} shows the result. Utilities is the most
persistently concentrated cluster in every period: once the market
is removed, the rate-driven utilities co-move as a standing block.
On top of that baseline each crisis lights up its own sector.
Technology concentration rises through 1999--2001 and collapses
after the dot-com bust. Energy is the dominant residual cluster of
the 2003--2012 period, already high in the 2005--2008 oil run,
with financials rising as a second mode into and through the 2008
crisis. Energy flips from cold before 2020 to a leading sector in
the Covid crash and the 2021 reflation. The market-removed leading
eigenvector thus supplies both the timing of a regime shift,
through the rotation and commutator of
Section~\ref{subsec:rotation}, and its identity, through the
sector composition, where a single eigenvalue would give neither.

\subsection{Name-level attribution: who moves, and who moves first}
\label{subsec:names}

The sector statement resolves to individual names. We track the full per-name
loading trajectory of the leading market-removed eigenvector at the two-year
lookback, aligning the sign of $v_1(t)$ in time so the trajectory is continuous,
take the pre-crisis anchor and the peak-rotation date where the eigenvector is
farthest from the anchor, and rank names by their loading change. Ordering the
top movers by when each crosses the halfway point of its shift, relative to
onset, shows the sequence in which the rotation happens.

\begin{figure}[htbp]
\centering
\includegraphics[width=0.98\textwidth]{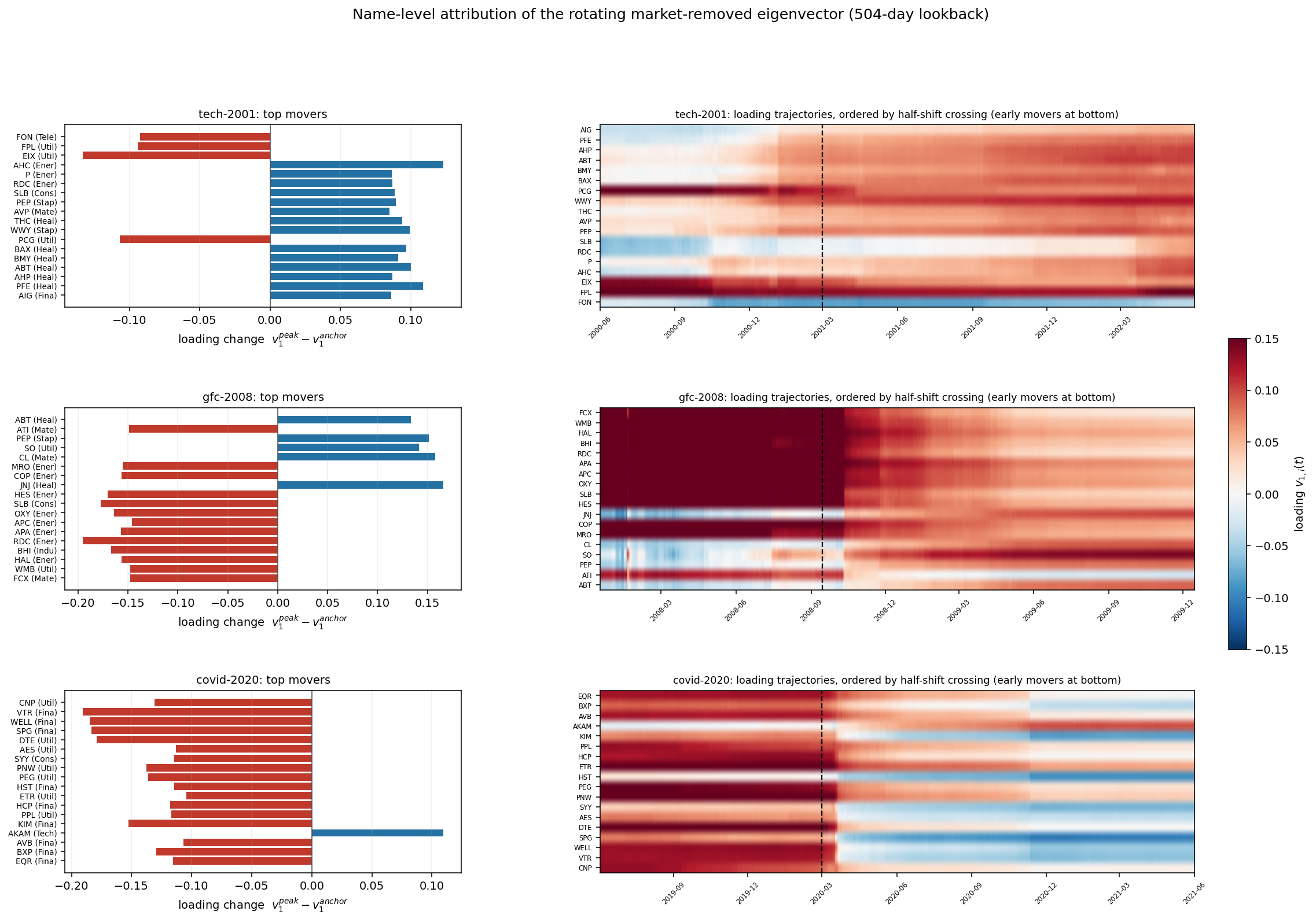}
\caption{Name-level attribution of the rotating market-removed eigenvector,
two-year lookback. Left: top movers by loading change from the pre-crisis
anchor to the peak-rotation date, blue for increasing and red for decreasing
loading. Right: loading trajectories of those names, ordered by the timing of
their half-shift crossing, with onset dashed. The 2020 movers change together
at onset, the 2001 movers are staggered with early movers, and the 2008 energy
cluster reorganizes gradually.}
\label{fig:names}
\end{figure}

The movers are the specific protagonists of each crisis
(Figure~\ref{fig:names}). In 2008 the largest loading moves are the energy
names, the oil majors, offshore drillers, and oilfield-service firms such as
Halliburton, Occidental, ConocoPhillips, Apache, and Schlumberger, which formed
the dominant residual cluster of the oil supercycle and disperse through the
crisis, while healthcare names move the other way. In 2020 the entire
rate-sensitive block moves together: the real estate investment trusts,
Simon Property, Ventas, Welltower, Host Hotels, Boston Properties, and the
utilities collapse from strongly positive loadings toward zero, with a software
name moving the other way. At the sector level the eigenvector rotates out of
this rate-sensitive complex and into energy, which takes over as the leading
residual cluster through the crash and the 2021 reflation
(Figure~\ref{fig:sectors}), so the yield-complex movers and the energy
concentration are the two ends of one rotation. In 2001 the movers are defensive, healthcare and
consumer staples gaining loading while utilities and telecom fall, the
name-level counter-rotation to the technology cluster whose
concentration peaks and collapses across the bust
(Figure~\ref{fig:sectors}).

The ordering answers who moves first, and it matches the forecasting result of
Section~\ref{subsec:earlywarning}. For Covid the top movers all cross their
half-shift within about two weeks of onset, a synchronized break with no early
movers, which is the exogenous-shock signature. For 2001 there are genuine early
movers, utilities and telecom shifting months ahead, consistent with the slow
dispersed unwind. For 2008 the energy cluster reorganizes gradually across
2008--2010, so at the two-year lookback the half-shift dates fall around and
after the trough, and a shorter lookback would place them earlier. The
synchronized versus staggered contrast across the trajectories is the name-level
face of why the 2008 crisis was forecastable from the correlation structure and
the 2020 crash was not.

\subsection{Early warning: which crises are forecastable}
\label{subsec:earlywarning}

The observables above describe a crisis as it unfolds. A separate question is
whether any of them warns of one in advance. We test this with a causal
criterion: each 126-day signal at date $t$ uses only the trailing window, and
the label is a forward peak-to-trough drawdown of the equal-weight market
beyond a threshold within the next 21 or 63 trading days. We score each signal
by the area under the ROC curve, pooled over the three periods and broken out
by period, using the rising mean correlation, the market-factor share, the
negative participation ratio, the one-month rotation rate, and the commutator
against the pre-crisis anchor.

\begin{figure}[htbp]
\centering
\includegraphics[width=0.82\textwidth]{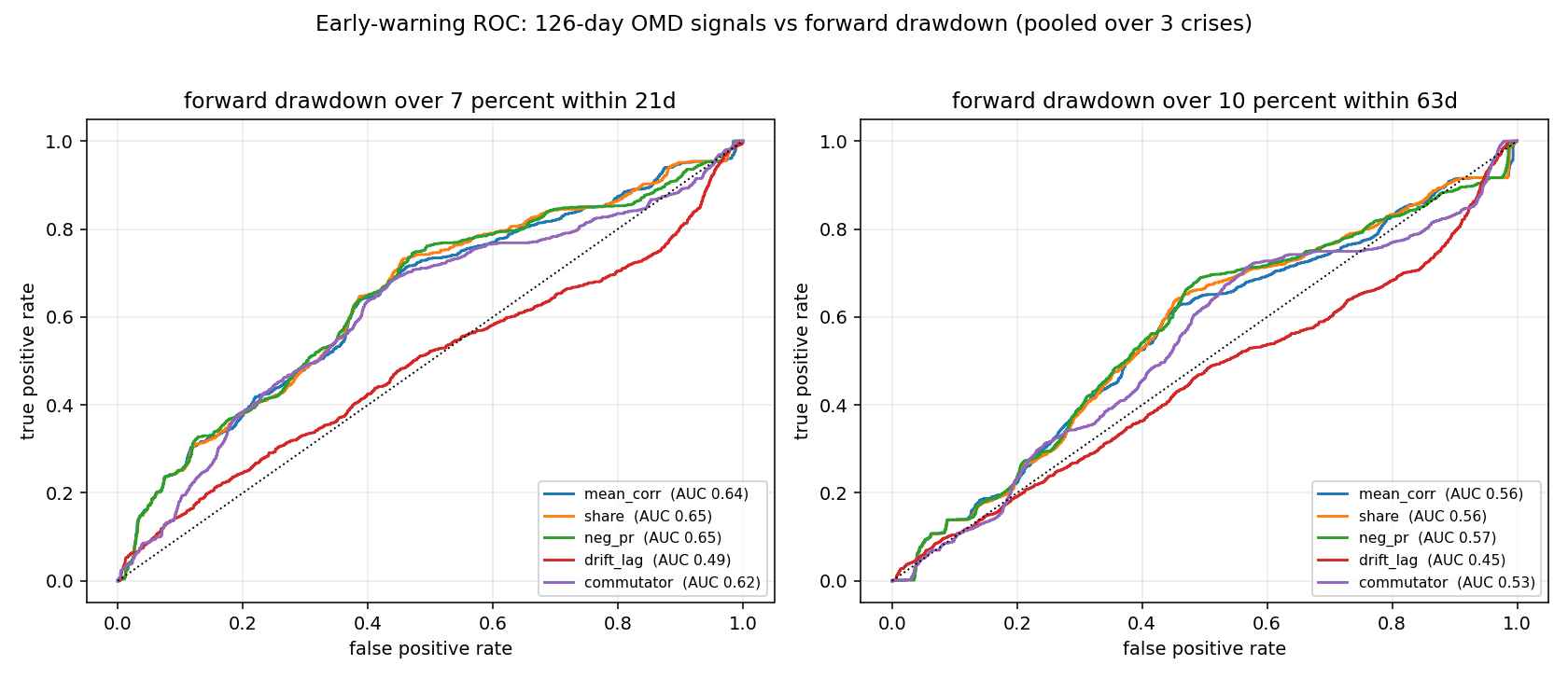}
\caption{Early-warning ROC for the 126-day signals against a forward market
drawdown, pooled over the three crises. Left: drawdown over seven percent
within 21 days. Right: drawdown over ten percent within 63 days. The
fragility levels lead modestly in the pool and strongly for the 2008 crisis
alone; the rotation rate is coincident. Signals use only trailing data, so the
test is causal.}
\label{fig:roc}
\end{figure}

The answer is regime-dependent, and that is the finding. The
correlation-fragility signals, rising mean correlation, rising market-factor
share, and falling participation ratio, give genuine early warning only for the
2008 crisis, which built up endogenously over a year: the per-period AUC is
about $0.72$ at the 63-day horizon, and the fragility crosses its
eightieth-percentile level months before the trough. The same signals have no
skill for the 2020 Covid crash (AUC about $0.49$) or the 2001 dot-com unwind
(AUC about $0.46$). This is the correct behavior rather than a failure. An
endogenous fragility measure can forecast a crisis that grows out of the
correlation structure itself, and cannot forecast an exogenous shock such as a
pandemic, or a dispersed decorrelated unwind. The pooled AUC, about $0.65$ at
21 days and $0.56$ at 63 days (Figure~\ref{fig:roc}), averages these regimes
and understates the per-period contrast. The eigenvector rotation rate is
coincident rather than leading, with an AUC near $0.49$: it spikes as the
eigenbasis reorganizes, which happens during the crash and not before.

\subsection{A geometric view: MDS embedding of the distance matrix}
\label{subsec:mds}

The spectral diagnostics summarize the distance matrix $M(t)$ through a
handful of numbers. A direct geometric view places the names in space.
Classical multidimensional scaling embeds $M(t)$ in three dimensions, one
point per name, so that the Euclidean distances of the embedding reproduce
the angular distances of $M(t)$ as closely as a three-dimensional picture
allows. We keep the off-subspace residual of each name as a small radial
puff, so the picture also carries the higher-rank content that three
coordinates drop. Sliding the sharp 126-day window through each crisis
turns the sequence of matrices into a moving cloud, aligned frame to frame
by a rotation so it does not flip at random. The embedding is a
visualization here, separate from the I-BBS spectral reading of $M(t)$.

\begin{figure}[htbp]
\centering
\includegraphics[width=\textwidth]{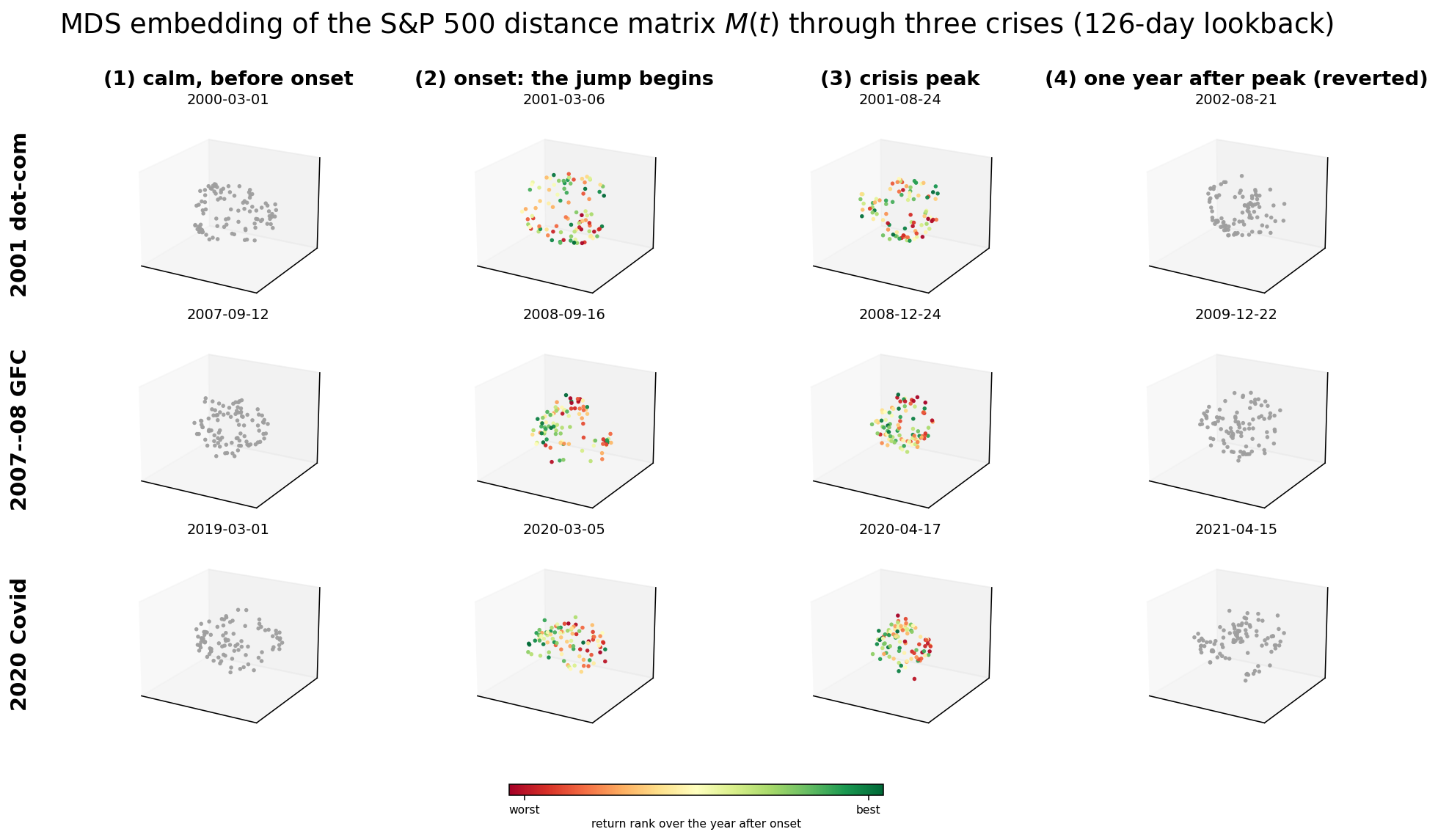}
\caption{MDS embedding of the distance matrix $M(t)$ through the three
crises, 126-day lookback. Rows: the three periods. Columns: a calm baseline,
the onset, the crisis peak, and one year after. Each name is a point, grey
outside the crisis and, from the onset to three months past the peak, shaded
by its return over the following year (green best, red worst).}
\label{fig:mds}
\end{figure}

Figure~\ref{fig:mds} shows four snapshots of the cloud for each crisis, read
left to right as a calm baseline a year before onset, the onset, the crisis
peak, and one year past the peak. The colour marks the crisis. Every name is
grey in calm times. At the onset the colour kicks in, each name shaded by its
total return over the following year and ranked across the universe, so green
is the best performers and red the worst. Three months after the peak the
colour reverts to grey, marking the return to normal conditions. Animated
versions of these embeddings, over the full period and around each onset,
accompany the paper.\footnote{The animated MDS embeddings, as
single-period videos and a combined multi-panel animation of the three
crises, are available in the paper's git repository,
\url{https://github.com/ighalp/omd_finance}.}

The geometry tracks the spectral story. In calm times the cloud is loose and
roughly isotropic. At the 2008 and 2020 onsets it contracts as the
correlations rise and the effective dimension collapses, most sharply for
Covid, where the cloud shrinks to a dense knot at the April 2020 peak. The
2001 dot-com period barely contracts, the dispersed decorrelated unwind that
the participation ratio already flagged, so its cloud stays spread. One year
past each peak the cloud has re-expanded to its calm size and the colour has
faded, the market back to normal. The picture is the spectral collapse seen
directly: a rise in correlation is a contraction of the name cloud.

\section{Stock dynamics in the ranking spaces}
\label{sec:markov}

The distance-matrix observable of the previous sections reads the
cross section through its correlations. A complementary reading
orders the names and follows the ranking. We use two orderings. The
first ranks the names by their rolling mean return over a short
window $\tau_{\mathrm{MC}}$,
\begin{equation}
\label{eq:rollret}
\bar r_i(t) = \frac{1}{\tau_{\mathrm{MC}}}
\sum_{s=t-\tau_{\mathrm{MC}}+1}^{t} r_i(s),
\end{equation}
a performance ordering. The second ranks them by their rolling
volatility over the same window, a risk ordering. In either case
rank $1$ is the leader, the highest mean return or the highest
volatility, and rank $N$ the laggard, and the rank of a name is a
discrete state whose evolution is a Markov chain on the
corresponding ranking space.

Both orderings are market-neutral by construction. A common move,
added to every return or scaling every volatility, leaves the order
unchanged, so each ranking discards the market factor that dominates
the raw correlation and keeps the relative, cross-sectional
dynamics. The one-step dynamics of each ordering are carried by a
time-dependent transition matrix,
\begin{equation}
\label{eq:Pmc}
P^{X}_{ab}(t) = \mathrm{Prob}\big(\text{rank } b \text{ at }
t+\Delta t \mid \text{rank } a \text{ at } t\big),
\qquad X \in \{R, V\},
\end{equation}
with $R$ the return ordering and $V$ the volatility ordering, each a
fixed-size $N\times N$ observable estimated by pooling the rank
transitions of all names over a window, with $\Delta t = 1$ day.
Because each rank is occupied by exactly one name on every date, the
marginal over ranks is uniform and the content lies entirely in the
transition structure. The band of $P^{X}$ around the diagonal
measures how sticky a rank is, its off-diagonal mass how often names
migrate, and the modulus of its second eigenvalue $|\lambda_2|$ sets
the mixing time $-1/\log|\lambda_2|$ over which an initial ranking is
forgotten. The two chains are read side by side throughout this
section, the return chain $P^{R}$ for relative performance and the
volatility chain $P^{V}$ for relative risk.

We treat these rankings as Markov chains only as a first
approximation, and do not claim that the dynamics on the ranking
spaces are memoryless. The transitions carry higher-order memory,
and the systematic and idiosyncratic factors specific to each firm
modulate how a name moves between ranks. Both effects can be built
in by driving the chain with exogenous covariates, such as firm
characteristics or a lagged state, which would add realism to the
dynamics on these spaces. In this paper we take the simplest view,
reading each one-step transition matrix as a Markov chain and
neglecting the covariates at leading order.

Four sources drive the migrations, and the ranking separates them.
The market factor drops out. A sector move carries the sector's
members up or down together, so it appears as a coherent block
migration and as a positive correlation between the one-step rank
changes of same-sector names. A firm-specific move shifts one name
against a still background. Pure noise reshuffles ranks at a rate
set only by the overlap of successive $\tau_{\mathrm{MC}}$-day
windows, which we read from an independent-return null that
preserves $N$, the window, and the sample length. The excess
persistence of the empirical chain over this null measures the
systematic and idiosyncratic relative-performance structure, and
the block coherence of the migrations measures the sector part.

The construction sits between two strands of the literature.
Relative and peer-based measures are standard for security selection
in a one-step, cross-sectional setting, where a name is compared
against its peers and the shared factor exposure cancels by
differencing. The relative-alpha measure of Jackwerth and Slavutskaya
\cite{jackwerth2018relative} nets a fund against its close peers, and
the peer cohorts of Forsberg, Gallagher, and Warren
\cite{forsberg2021peer} group funds by their return correlations and
score each against its group, both market-neutral by construction in
the same way our ranking is, and both used to select rather than to
follow the whole cross section over time. The dynamics of the
whole ordering, on the other hand, are the object of stochastic
portfolio theory and its rank-based diffusions
\cite{fernholz2002,banner2005}, which center on the stationary
capital-distribution curve and the collision dynamics of ranked
weights. Our reading is the missing dynamic complement: the whole
ordering as a fixed-size transition matrix $P(t)$, in place of the
distance matrix $M(t)$, whose spectrum, entropy production, and
transfer entropy we track through time rather than its stationary
shape.

\begin{figure}[htbp]
\centering
\includegraphics[width=0.98\textwidth]{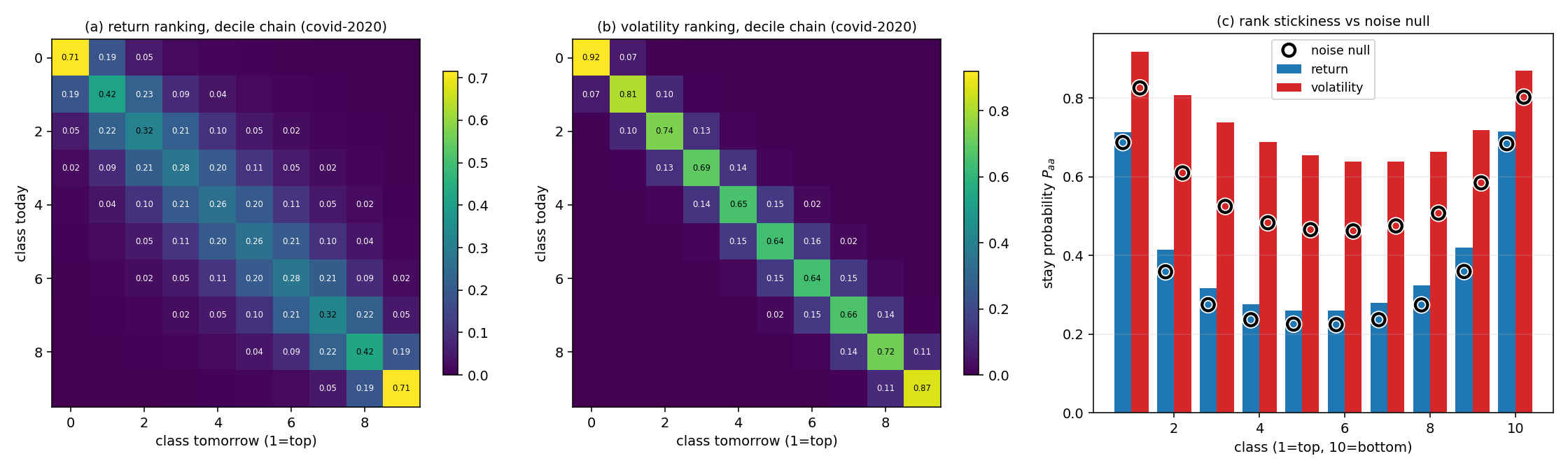}
\caption{Decile rank transition matrices for the Covid universe, at
$\tau_{\mathrm{MC}}=10$ days for the return chain and $21$ days for
the volatility chain. (a) The return chain, a broad band with the
extreme deciles holding at $0.71$ and the middle deciles spreading to
neighbours. (b) The volatility chain, a tight diagonal holding
between $0.64$ and $0.92$. (c) The per-decile stay probability of the
two chains against their independent-return noise nulls (open
circles), exceeded in every decile.}
\label{fig:markovP}
\end{figure}

Figure~\ref{fig:markovP} shows the two decile transition matrices
for the Covid universe. The return chain is a broad band. The
extreme deciles are the stickiest, the top and bottom holding their
class from one day to the next with probability $0.71$, while the
middle deciles hold theirs between $0.26$ and $0.42$ and spread to
their neighbours. The volatility chain is a much tighter diagonal,
every decile holding with probability between $0.64$ and $0.92$, so
a name keeps its risk rank far longer than its performance rank.
Both chains exceed their independent-return nulls in every decile,
the return chain most in the middle of the distribution and the
volatility chain across the board, so each ordering carries a
genuine persistence beyond the mechanical stickiness that the
overlapping rolling windows produce on their own.

\begin{figure}[htbp]
\centering
\includegraphics[width=0.78\textwidth]{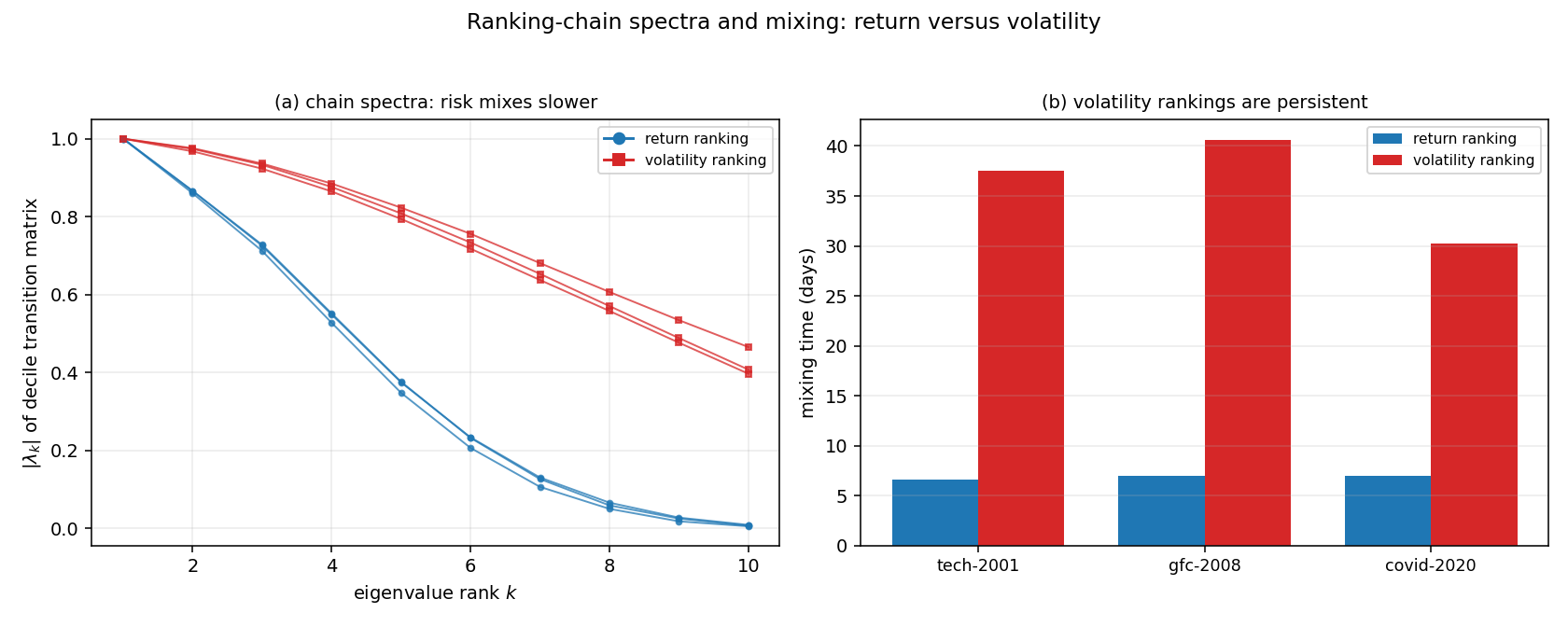}
\caption{Spectra and mixing of the two ranking chains. (a) Moduli of
the decile transition-matrix eigenvalues ranked by magnitude, with the
return chain (blue) and the volatility chain (red) each drawn for all
three periods as three near-coincident curves. The volatility spectrum
stays high in every period, so the volatility chain mixes slowly,
while the return spectrum decays quickly. (b) The corresponding mixing
times, one bar pair per period, about seven days for the return
ranking and thirty to forty for the volatility ranking.}
\label{fig:markovspec}
\end{figure}

The spectra make the contrast quantitative
(Figure~\ref{fig:markovspec}). The return transition matrix has its
second eigenvalue near $0.86$ and the rest falling to zero within a
few modes, so an initial performance ranking is forgotten in about
seven days. The volatility eigenvalues stay high across the whole
spectrum, the second near $0.97$, and the mixing time is thirty to
forty days, five to six times longer. This is the spectral signature
of volatility clustering. Risk is persistent and predictable, a
turbulent name stays turbulent for weeks, while relative performance
churns fast, a leader today need not be one tomorrow. The two chains,
built from the same names by the same construction, sit at opposite
ends of the persistence scale.

The sector factor is visible directly in the migrations. In the
return ranking the one-step rank changes of same-sector names
correlate at $0.08$ to $0.10$ across the three periods, against
near-zero for the full set, so sector members move up and down the
performance ordering together. This block coherence strengthens from
the 2001 period to the Covid period, tracking the tightening sector
structure of the correlation analysis. In the volatility ranking the
same within-sector correlation is weaker, $0.03$ to $0.05$, because a
name's risk rank is set more by its own idiosyncratic turbulence than
by a sector-wide move.

\begin{figure}[htbp]
\centering
\includegraphics[width=0.92\textwidth]{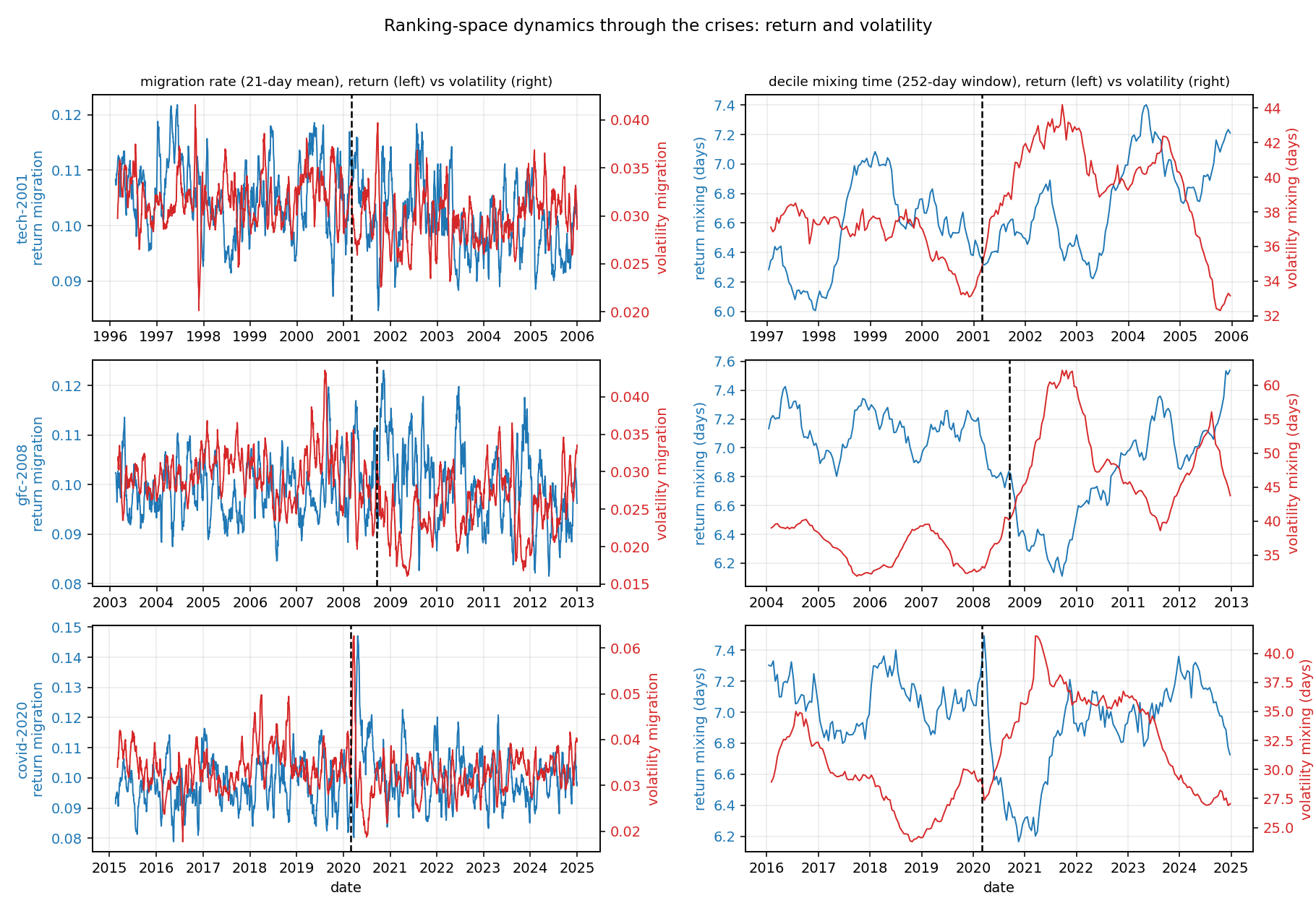}
\caption{Ranking-space dynamics through the crises, return and
volatility. In both columns the return chain (blue) is on the left
axis and the volatility chain (red) on the right, each on its own
scale, since the two chains move over very different ranges. Left: the
migration rate, the mean one-step class change per name over $N$, as a
21-day mean, spiking at each onset for both chains. Right: the mixing
time of the decile chain on a 252-day window. The return mixing time
stays near a week and dips as an onset passes, while the volatility
mixing time is an order of magnitude longer and rises and stays
elevated after each onset. The dashed black line marks the crisis
onset used throughout the paper, 1 March 2001 for the dot-com period,
the Lehman collapse on 15 September 2008 for the financial crisis, and
1 March 2020 for the Covid period.}
\label{fig:markovdyn}
\end{figure}

The ranking dynamics carry a sharp crisis signature
(Figure~\ref{fig:markovdyn}). At each onset the migration rate of
both chains spikes, most violently for Covid, where the mean one-step
change jumps by about half as the ordering reshuffles in the crash.
The two chains then part in their mixing. The return mixing time
drops slightly and sits at or below its pre-crisis level, so the
turbulent aftermath churns the performance ordering a touch faster.
The volatility mixing time instead rises sharply and stays elevated
for a year or more, climbing toward sixty days after the 2008 crisis,
so a crash makes the risk ordering more persistent even as it makes
the performance ordering less so. This is complementary to the
correlation reading. A crash raises the average correlation, which
co-moves the names, and at the same time widens the dispersion of
their magnitudes, which churns their performance order and locks in
their relative risk.

\subsection{Entropy production of the ranking dynamics}
\label{subsec:entropy}

The rank transition matrix invites a further reading from
stochastic thermodynamics. A stochastic process is time-reversible
only if its forward and backward paths are equally likely, and the
failure of that balance is measured by the entropy production, the
Kullback-Leibler divergence between the forward and backward path
probabilities \cite{seifert2005,seifert2012}. A recent application
of this stochastic-thermodynamic reading to financial markets,
computing entropy production and the second law in the high-frequency
single-security setting, is given by Georgiev \cite{georgiev2025}. For a one-step
transition with forward matrix $P$, time-reversed matrix $\tilde P$, and
initial and final marginals $p$ and $p'$, the total entropy
production is
\begin{equation}
\label{eq:sdot}
\sigma = \sum_{ab} p_a P_{ab}
\log \frac{p_a P_{ab}}{p'_b \tilde P_{ba}}
= \underbrace{\sum_{ab} p_a P_{ab}
\log \frac{P_{ab}}{\tilde P_{ba}}}_{\sigma_{\mathrm{med}}}
+ \underbrace{H[p'] - H[p]}_{\sigma_{\mathrm{sys}}},
\end{equation}
the sum of a medium (irreversibility) part and a system entropy
change \cite{seifert2005,halperin2026entropy}. In an
information-engine setting the backward matrix $\tilde P$ is often
hidden, and only a lower bound on $\sigma$ is available
\cite{halperin2026entropy}. Here the whole trajectory is observed,
so both the forward and the backward transition matrices are
estimated directly from the rank data.

For the ranking chain the class marginal is fixed by the permutation
structure, a fixed number of names in each class at every date, so it
is identical before and after a step and the system term vanishes
exactly. Estimating the backward transitions from the reversed-time
rank data transposes the joint, so the entropy production over an
interval $\Delta t$ is the divergence between the forward joint
rank-transition distribution $\mu_{ab}(\Delta t) =
\mathrm{Prob}(\text{rank } a \text{ at } t,
\ \text{rank } b \text{ at } t+\Delta t)$ and
its time reverse $\mu_{ba}(\Delta t)$,
\begin{equation}
\label{eq:epmkt}
\sigma(\Delta t) = \sum_{ab} \mu_{ab}(\Delta t)
\log \frac{\mu_{ab}(\Delta t)}{\mu_{ba}(\Delta t)},
\end{equation}
which is non-negative and vanishes exactly when $\mu$ is symmetric,
that is when the rank dynamics satisfy detailed balance. It measures
the time-irreversibility of the relative-performance dynamics. Any
nonzero $\sigma$ is a genuine breaking of time-reversal symmetry, so
the relevant question is not whether it vanishes but how large it is.
A small $\sigma$ marks dynamics that are only weakly non-equilibrium,
close to the reversible, equilibrium limit rather than far from it. A
symmetric diffusion of ranks produces none. A directed flow, such as
the asymmetry between a fast crash and a slow recovery or a
momentum-then-reversion cycle, produces a positive current
$J_{ab} = \mu_{ab} - \mu_{ba}$ and a positive $\sigma$.

\begin{figure}[htbp]
\centering
\includegraphics[width=0.98\textwidth]{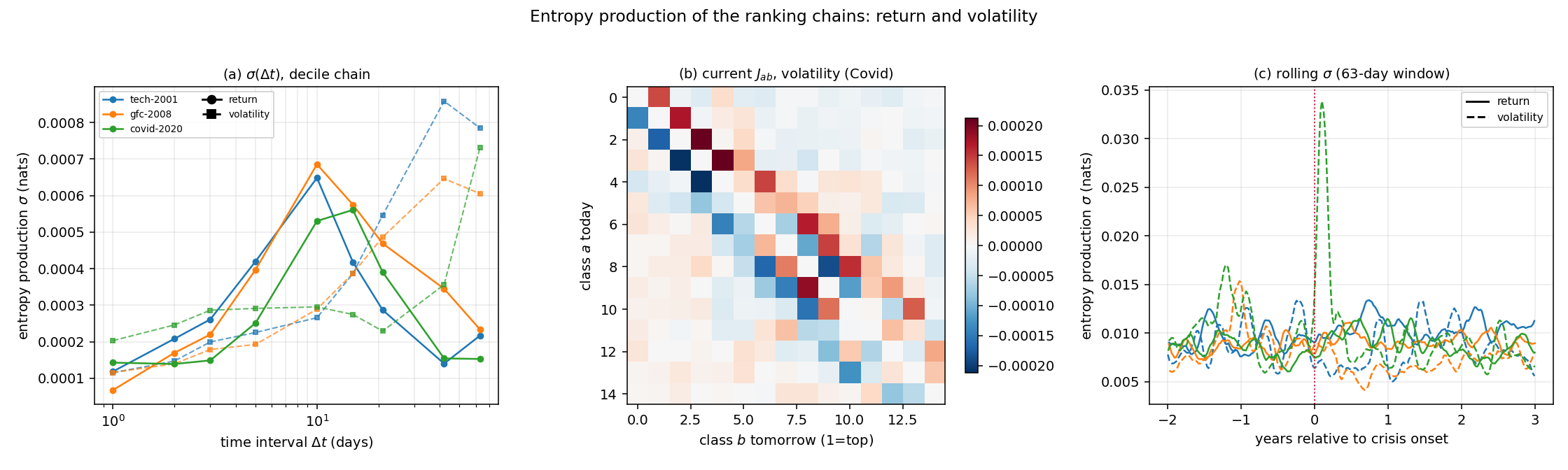}
\caption{Entropy production of the ranking-space Markov chains,
return and volatility. (a) $\sigma(\Delta t)$ from
Eq.~\eqref{eq:epmkt} against the time interval, return (solid) and
volatility (dashed), each peaking near its own rolling-window length.
(b) The probability current $J_{ab}=\mu_{ab}-\mu_{ba}$ of the
volatility chain for the Covid universe, the antisymmetric
circulating flow. (c) Rolling entropy production on a 63-day window
against time-to-onset, return (solid) and volatility (dashed); the
volatility chain lifts far more, spiking at the Covid onset.}
\label{fig:entropy}
\end{figure}

Figure~\ref{fig:entropy} reports the estimate for both chains. Both
are close to detailed balance: the decile entropy production is of
order $10^{-3}$ nats per step, small because rank motion is nearly
one-dimensional, and a birth-and-death chain on a line is reversible
in steady state. What structure there is has a clear timescale.
Panel (a) shows that $\sigma(\Delta t)$ is small at one day, rises to
a maximum near the rolling-window length, ten days for the return
chain and about twenty for the volatility chain, and decays for
longer intervals. Shorter intervals share most of their averaging
window and move the ranks reversibly, while longer intervals let the
chain mix back toward the uniform distribution. Panel (b) shows the
probability current $J_{ab}$ of the volatility chain for the Covid
universe, an antisymmetric pattern concentrated near the diagonal and
at the extreme classes, the signature of the systematic circulation
that carries its arrow of time. Panel (c) tracks the rolling entropy
production of both chains through the crises. Both lift in raw terms
after each onset, and the volatility chain lifts far more, most sharply
at the Covid onset, as the fast drops and slow recoveries of the crash
raise the raw asymmetry above its calm-time level. Whether that raw
lift is a genuine arrow or a finite-sample response is what the
surrogate test settles below.

\begin{figure}[htbp]
\centering
\includegraphics[width=0.98\textwidth]{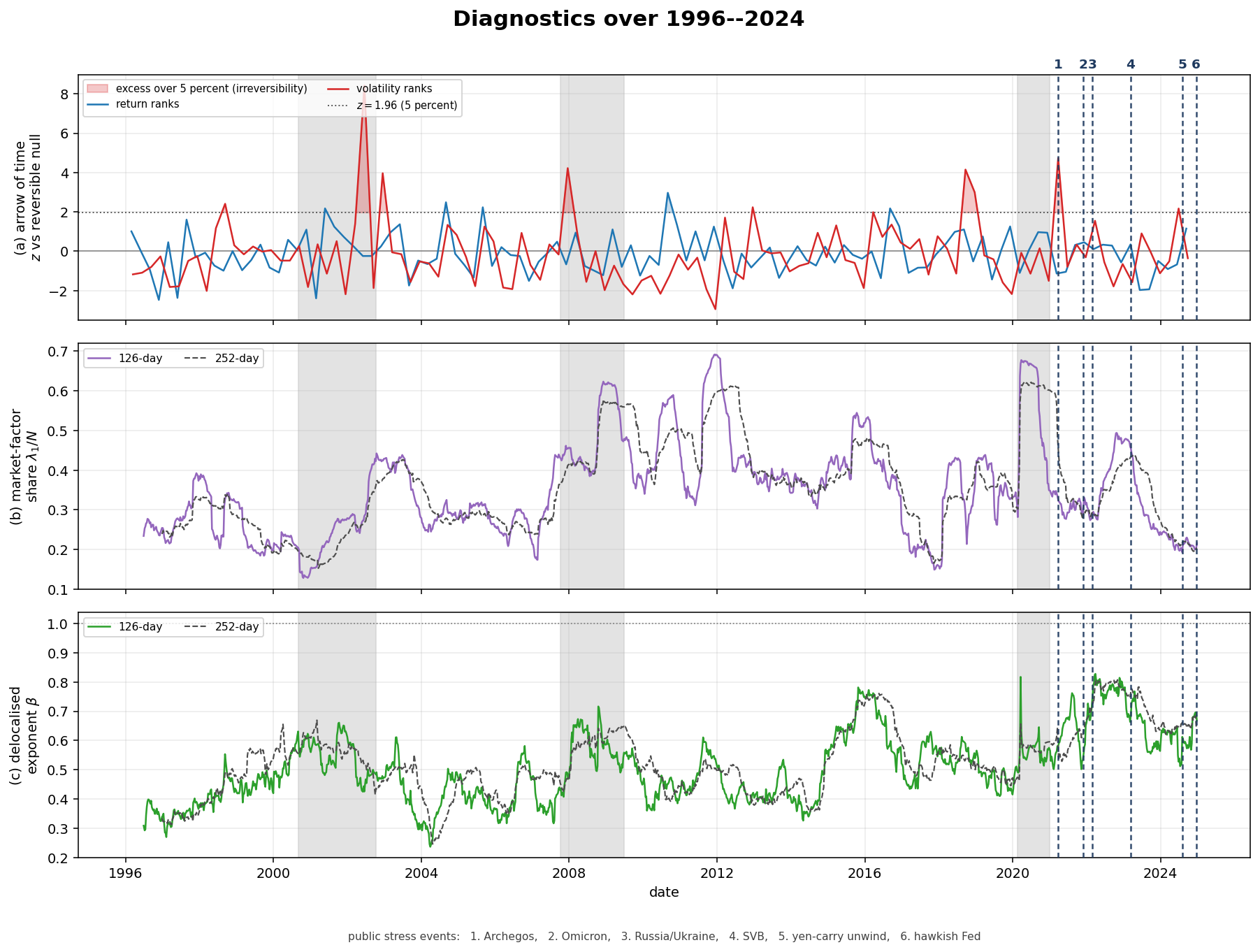}
\caption{Diagnostics over 1996--2024, resolved in time, against the
market-crisis regimes (shaded) and public market-stress events
(numbered dotted lines: 1 Archegos, 2 Omicron, 3 Russia/Ukraine, 4
SVB, 5 yen-carry unwind, 6 hawkish Fed). (a) The arrow of time, the
entropy-production $z$-score of the decile chain against the reversible
detailed-balance null on 63-day blocks for one fixed whole-span
universe, for the return ranking (blue) and the volatility ranking
(red), with the shaded excess above the five-percent line marking the
periods of significant irreversibility. The return arrow rarely clears
the five-percent line, while the
volatility arrow flares at market stress, strongest at the 2002 dot-com
bottom ($z\approx 8$) and the 2007--08 crisis ($z\approx 4$). Among
the marked single-session events only Archegos leaves a clear arrow.
(b) The market-factor share $\lambda_1/N$ and (c) the delocalised
exponent $\beta$ of $M(t)$, computed on the hundred largest-cap names
and shown at a 126-day (solid) and a 252-day (dashed) rolling window.
Both rise at every crisis regime and around the SVB banking stress
where the arrow is flat, while Archegos moves the arrow but not the
share. The two windows agree, the longer one smoothing the shorter,
and both read regimes rather than single sessions.}
\label{fig:volarrow}
\end{figure}

The non-negativity of the entropy production is automatic.
Equation~\eqref{eq:epmkt} is a Kullback-Leibler divergence, so
$\sigma\ge 0$ by construction, the one-step form of the second law,
and it is exactly what makes the forward and backward paths
statistically distinguishable, the operational content of the arrow
of time \cite{parrondo2009}. A positive value therefore does not by
itself establish an arrow. The plug-in estimator is biased upward at
finite sample, and the question is whether the observed $\sigma$
exceeds what a reversible process of the same size would give. We
test this by comparing each block of the chain against a
detailed-balance surrogate, a reversible chain matched to the block's
marginals and sample size, and reading the excess of the observed
entropy production over the surrogate. Figure~\ref{fig:volarrow}(a)
reports this excess as a $z$-score across 1996--2024 for one fixed
whole-span universe.

The answer is that the return-ranking dynamics are close to detailed
balance. Its $z$-score stays near zero across the span and never
drifts upward, so no irreversibility accumulates. A direct test is
decisive. Comparing the entropy production of the one-year window
after each onset against the reversible surrogate, no crisis is
significantly above the null, and the 2008 and 2020 windows sit
slightly below it, so the year that follows a crash restores
reversibility in the return ranking.
Pooling the whole 1996--2024 span into a single estimate, over a
million transitions with negligible bias, is equally decisive for a
general arrow: the entropy production stays within statistical error
of the reversible null at every resolution, and slightly below it at
the decile level. In relative-performance space the market is
therefore close to time-reversible, near equilibrium.
This is consistent with what is known of time-reversal asymmetry in
finance, which is pronounced in the volatility, through the Zumbach
effect where past volatility forecasts future volatility more than
the reverse \cite{zumbach2009}, and weak at the level of the return
ordering that the ranking reads. The raw entropy production still
lifts after a crisis, as Figure~\ref{fig:entropy}(c) shows, but the
lift is largely the finite-sample and structural response rather than
a large excess over reversibility. It also lags the spectral jump by
several months, because the irreversible signature needs both the
fast crash and the slow recovery in the window before the forward and
backward transitions differ.

Comparing the two chains isolates the arrow of time, and resolving it
in time shows where it concentrates. Time-reversal asymmetry in markets is
known to reside in the volatility rather than the returns, so the
volatility chain should carry an arrow that the return chain does not.
Figure~\ref{fig:volarrow} tracks the entropy-production $z$-score of
both chains quarterly across 1996--2024 on one fixed universe, with
the market-stress episodes shaded. The return chain stays near the
reversible null throughout, rarely clearing the five-percent line and
with no systematic tie to a crisis. The volatility chain is reversible
in calm times, its baseline sitting slightly below zero, but carries a
weak, episodic arrow that flares at market stress.

The largest flares are the 2002 dot-com bottom, reaching $z\approx 8$,
and the 2007--08 financial crisis at $z\approx 4$, with further flares
at the 2018 selloff and through the 2020--21 volatility regime.
Testing the trace against a calendar of documented single-session
stress events, marked on the figure, sharpens what it measures. Of
those events only the March 2021 Archegos liquidation leaves a clear
arrow, at $z\approx 5$, with the August 2024 yen-carry unwind a
marginal one. The Omicron scare, the SVB failure, and the hawkish-Fed
repricing raise volatility sharply yet leave no arrow at all. The
reason is twofold. A one-session shock is diluted inside a quarterly
block, and a spike that falls and recovers symmetrically carries no
time asymmetry to begin with. The arrow responds to sustained,
directional reshuffling of the cross section, the forced multi-day
deleveraging of Archegos or the grinding drawdowns of the large
crises, rather than to volatility on its own.

The absolute entropy production stays small at every date, so even the
flares are only weakly non-equilibrium, yet they are genuine and they
sit where the stochastic-volatility literature places the market's
arrow of time. The March 2020 Covid crash is the informative
exception. Too brief for a quarterly block to resolve, its arrow does
not appear in the crash quarter itself but in the sustained
high-volatility regime that runs into 2021. A pooled estimate over a
whole multi-year period inherits this smearing, which is why the arrow
is better read from the time-resolved trace than from a single number
per crisis.

The same figure carries two of the distance-matrix spectral
diagnostics on the same axis, as a check on whether they move where the
arrow does not. Panels (b) and (c) track the market-factor share
$\lambda_1/N$ of $C(t)$ and the delocalised exponent $\beta$ of $M(t)$,
computed on the hundred largest-cap names and shown at both a half-year
(126-day) and a one-year (252-day) rolling window. Both rise at every
crisis regime, the share climbing from a calm level near $0.33$ to
$0.58$ in 2008 and $0.68$ at Covid and the exponent steepening from
near $0.49$ toward $0.80$, so a crash concentrates the cross section
and lifts the whole battery of diagnostics at once. They also separate
the single-session events by type. The share and the exponent both lift
around the SVB banking failure, where the arrow of time is flat, so the
spectral concentration reads a correlation-regime shift that the arrow
does not, whereas Archegos moves the arrow while leaving the share at
its calm level, a directional reshuffling without broad concentration.
The two windows agree on this structure, the longer one smoothing the
shorter, and at either window these are regime indicators that flag the
sustained stress around an event rather than the single session.

\begin{figure}[htbp]
\centering
\includegraphics[width=0.98\textwidth]{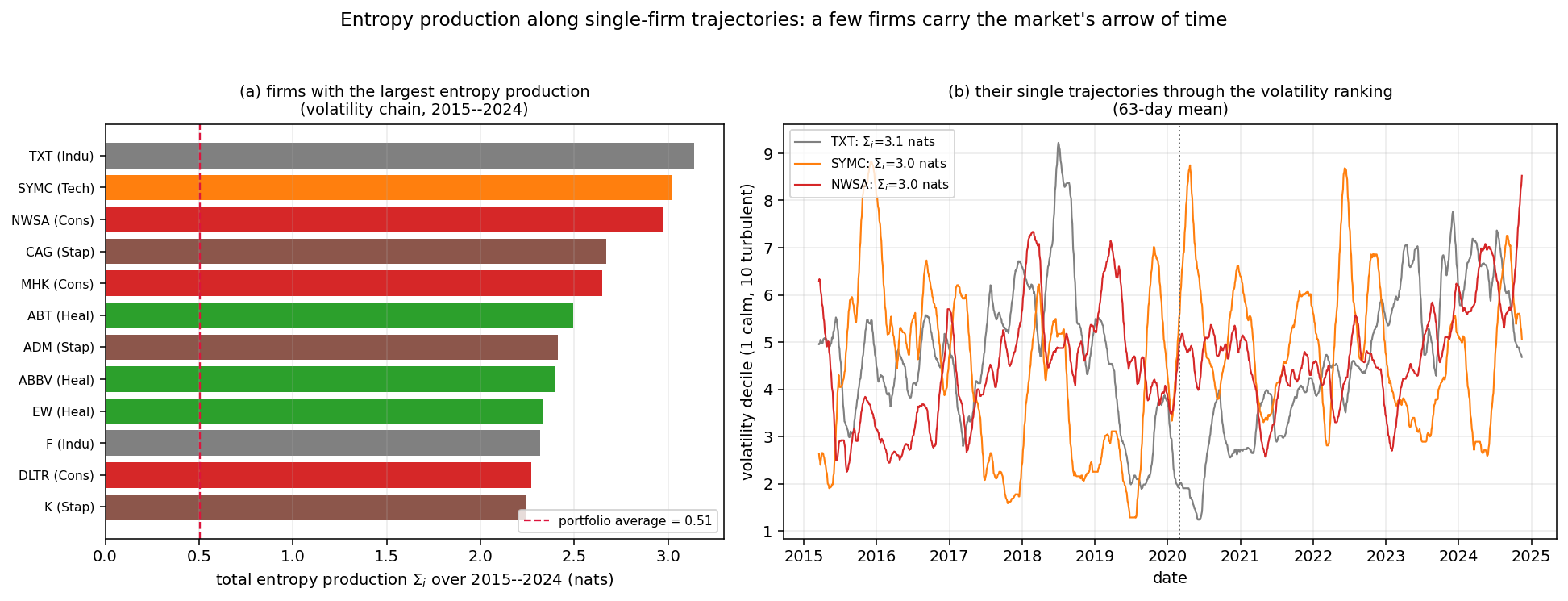}
\caption{Entropy production along single-firm trajectories, volatility
chain, Covid decade. (a) The twelve firms with the largest
single-trajectory entropy production $\Sigma_i=\sum_t F$, the
thermodynamic force accumulated along each firm's own path through the
volatility ranking over the decade, coloured by sector. The portfolio
average of $\Sigma_i$ (dashed, $0.5$ nats) is the pooled entropy
production integrated over the interval, and the top firms exceed it
five to six times. Dividing $\Sigma_i$ by the $2495$ daily steps gives
the per-step rate. (b) The volatility-decile trajectories of the top
three firms (63-day mean), the single paths whose entropy production
panel (a) reports, with the Covid onset dotted.}
\label{fig:firmentropy}
\end{figure}

The entropy production so far is a per-step portfolio average, but
stochastic thermodynamics also assigns an entropy production to each
single trajectory, here the path of one firm through the volatility
ranking over the whole interval. The thermodynamic force
$F_{ab}=\log(\mu_{ab}/\mu_{ba})$ is a property of the pooled chain, and
firm $i$'s entropy production $\Sigma_i=\sum_t F$ is the force
accumulated along its own trajectory, the total over the decade rather
than a rate. Dividing $\Sigma_i$ by the $2495$ daily steps returns a
per-step rate, and the portfolio average of that rate is the pooled
$\sigma$ exactly. Figure~\ref{fig:firmentropy} decomposes the
Covid-decade volatility arrow this way, and the distribution is very
uneven. A handful of firms accumulate about $3$ nats over the decade,
five to six times the portfolio-average $0.5$, so the market's small
arrow of time is not spread evenly but concentrated in a few names
whose volatility rank rides the systematic fast-up, slow-down
circulation of volatility clustering. The leaders in this interval are
Textron, Symantec, News Corp, Conagra, and Mohawk, a mix of
industrial, technology, and consumer names rather than a single
sector.

\subsection{Transfer entropy across sectors: who moves whom}
\label{subsec:transfer}

The entropy production is a single number for the whole chain. A
finer, directional question is whose rank movements predict whose.
It is answered by the transfer entropy \cite{schreiber2000}, the
reduction in uncertainty about one series' next state gained from a
second series' present, beyond the first series' own present,
\begin{equation}
\label{eq:te}
\mathrm{TE}(S\to T) = \sum p(t', t, s)
\log \frac{p(t' \mid t, s)}{p(t' \mid t)},
\end{equation}
where $t'$ is the next state of the target $T$, and $t$ and $s$ the
present states of $T$ and the source $S$. It vanishes when $S$
carries no information about the future of $T$ that $T$ does not
already carry. On continuous return series the transfer entropy
needs a nearest-neighbor estimator and a surrogate test
\cite{kraskov2004}. The ranking removes that difficulty. With the
states already the rank classes, Eq.~\eqref{eq:te} is an exact
plug-in histogram estimator, and only a shuffle of the source is
needed to subtract the finite-sample bias floor.

\begin{figure}[htbp]
\centering
\includegraphics[width=0.98\textwidth]{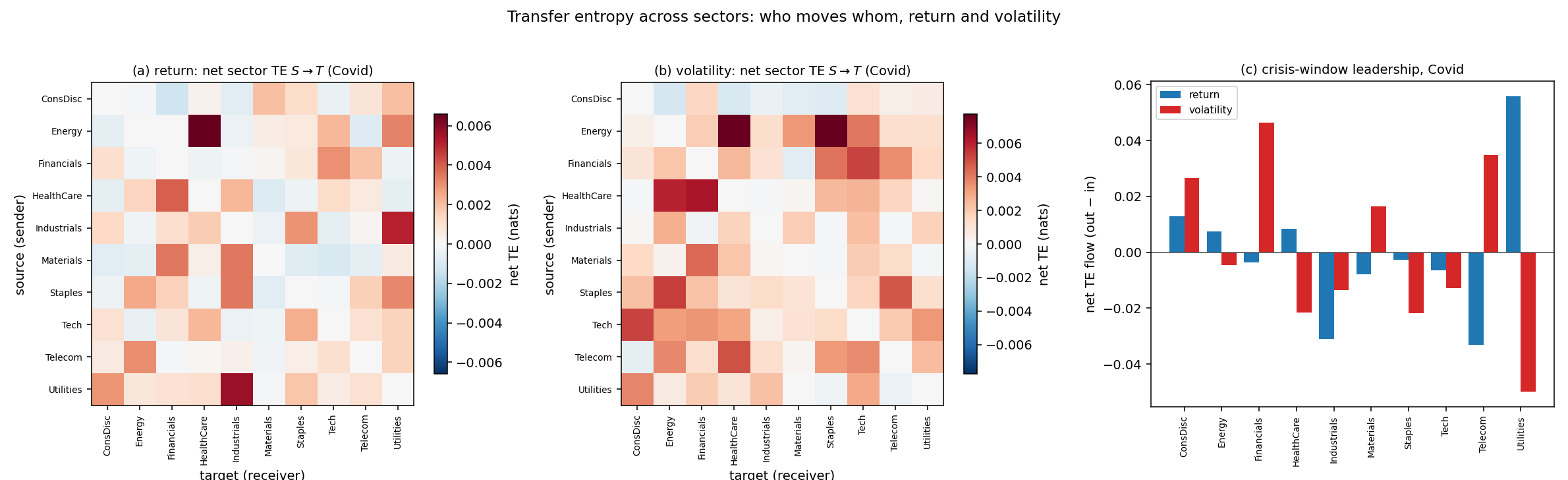}
\caption{Directed transfer entropy between sectors, each represented
by the tercile of its mean rank, surrogate-corrected, for the Covid
period. (a) The net sector transfer-entropy matrix $S\to T$ for the
return ranking. (b) The same for the volatility ranking. (c) The
crisis-window net flow, return against volatility: utilities leads
the return channel while financials lead the volatility channel.}
\label{fig:transfer}
\end{figure}

We coarse-grain each sector to the tercile of its mean rank, hot,
warm, or cold, and compute the surrogate-corrected transfer entropy
between every ordered pair of sectors, in both rankings. The net flow
of a sector, its total outgoing minus incoming transfer entropy,
ranks it as a net sender, a leader whose moves predict others, or a
net receiver, a follower. Figure~\ref{fig:transfer} shows the two
directed networks for the Covid period (panels a and b) and the
crisis-window leadership of both (panel c). In the return ranking the
crisis leadership is a flight to quality. Utilities is the dominant
net sender in the 2008 and Covid windows, with health care close
behind, and is among the leaders in 2001, so the rate-driven
defensive sectors set the pace of the performance ordering while the
cyclicals follow. This is the same utilities block that stands out as
the persistent residual cluster in the distance-matrix analysis,
identified here as the crisis-time source rather than only the most
coherent group.

The volatility ranking tells a different story about the same crises.
There the net senders are the financials, dominant in both the 2008
and the Covid windows, while utilities turns from the strongest
performance leader into a risk follower, one of the largest net
receivers. The two networks split the crisis into a performance
channel, where defensives lead a flight to quality, and a risk
channel, where the financial sector's turbulence propagates first. In
the 2001 dot-com period, which was not a banking crisis, financials
lead neither channel and the volatility leadership is diffuse. Reading
who moves whom in performance and who moves whom in risk gives two
directed networks that a single return series would merge into one.

\subsection{The lead-lag network at the name level}
\label{subsec:leadlag}

\begin{figure}[htbp]
\centering
\includegraphics[width=0.98\textwidth]{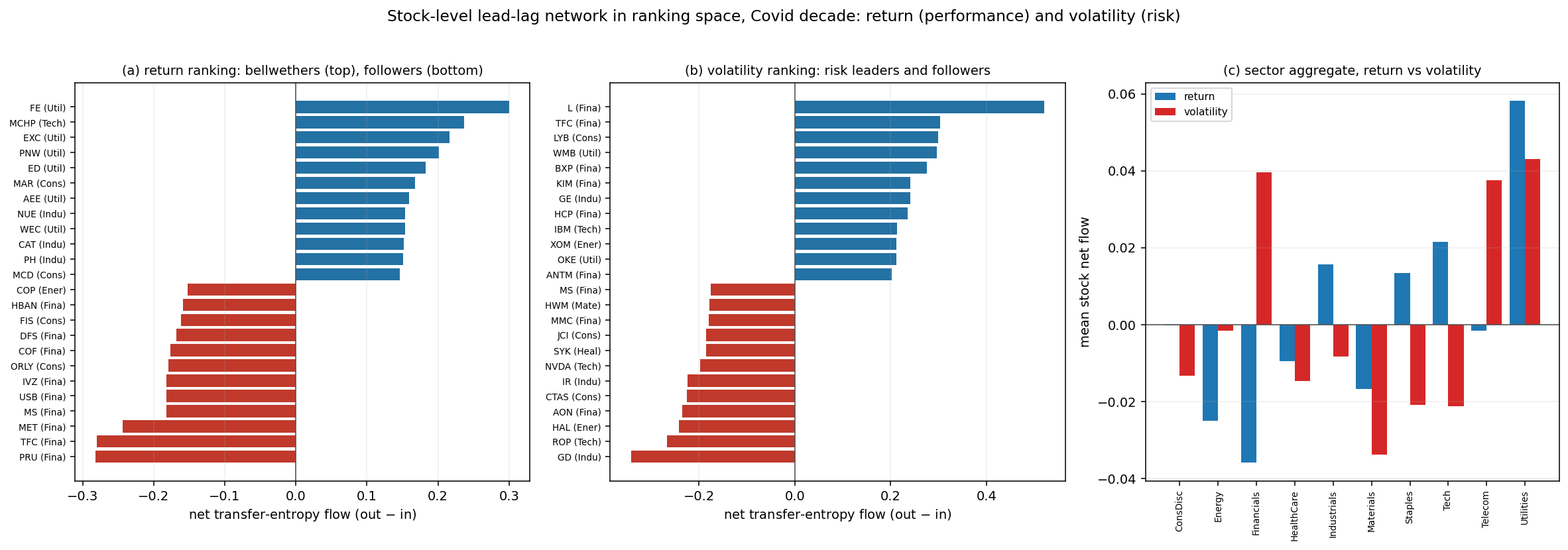}
\caption{Stock-level lead-lag network in ranking space, Covid decade.
(a) Return ranking: the net transfer-entropy flow per name, the top
twelve bellwethers and bottom twelve followers, blue for senders and
red for receivers, with utilities leading and financials following.
(b) Volatility ranking: the risk leaders and followers, where
financials and real estate lead. (c) The stock net flows aggregated
by sector, return against volatility, showing the role reversal
between the two channels.}
\label{fig:leadlag}
\end{figure}

The sector transfer-entropy network resolves to individual names, in
both rankings. Computing the discrete transfer entropy between every
pair of stocks in tercile rank space and ranking each by its net
flow, its outgoing minus incoming transfer entropy, gives a directed
lead-lag network whose senders are bellwethers and whose receivers
are followers (Figure~\ref{fig:leadlag}). In the return ranking for
the Covid decade the bellwethers are the utilities, FirstEnergy,
Exelon, Pinnacle West, Ameren, and their peers, whose rank moves
predict the rest, and the followers are the financials, Prudential,
Truist, MetLife, and Morgan Stanley. In the volatility ranking the
roles rotate. The risk leaders are the financials and real-estate
names, Loews, Truist, Boston Properties, and Kimco, whose turbulence
moves first, and the risk followers are the industrials and defensive
names, General Dynamics, Roper, and Aon. Truist is a clean instance
of the reversal, a follower in performance and a leader in risk.
Aggregating the stock net flows by sector (panel c) reproduces the
sector split, utilities leading the return channel and following the
risk channel while financials do the reverse. The name-level network
names the specific bellwethers behind each channel, the rate-driven
utilities that set the pace of relative performance and the financials
whose risk propagates first.

\subsection{Co-movement and dispersion}
\label{subsec:dispersion}

\begin{figure}[htbp]
\centering
\includegraphics[width=0.98\textwidth]{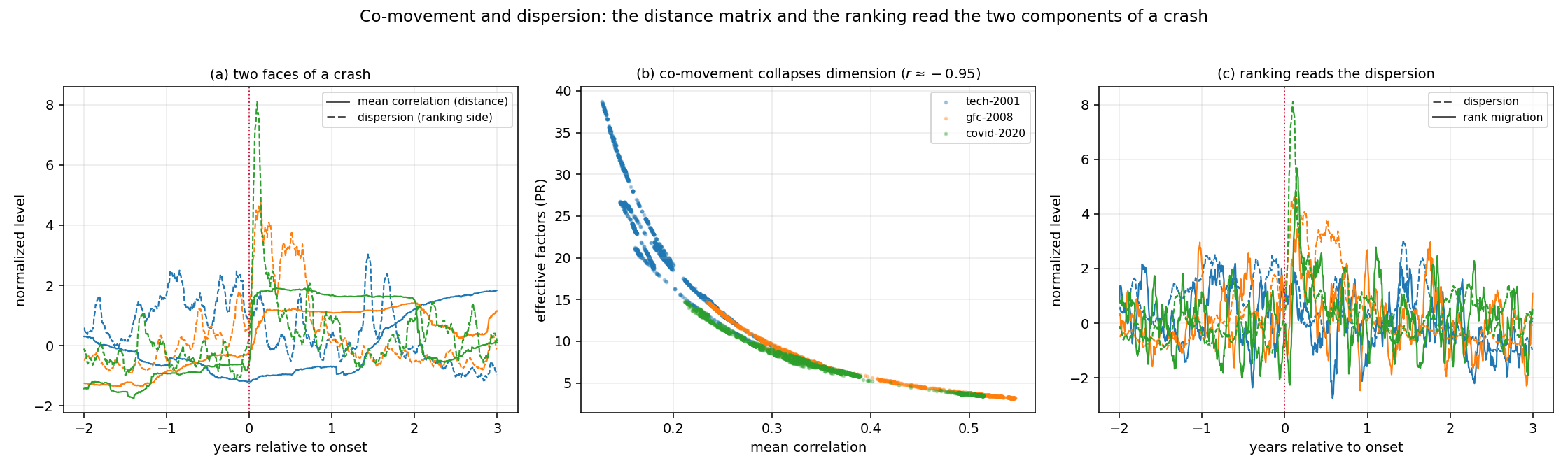}
\caption{Co-movement and dispersion, the two components of a crash.
(a) Aligned on onset, the mean correlation (co-movement, distance
matrix) and the cross-sectional dispersion (ranking chain) both rise
at the crash. (b) The participation ratio against the mean
correlation, a tight inverse curve shared by the three periods. (c)
The dispersion and the rank migration rate rise together through the
crash.}
\label{fig:dispersion}
\end{figure}

The two matrix views correspond to a decomposition of the daily
cross section. Each day the panel of returns is the sum of a market
move, shared by every name, and a residual spread around it, the
cross-sectional dispersion. The distance matrix is dominated by the
market factor and reads the systematic co-movement; the ranking chain
is market-neutral and reads the residual dispersion. A crash
amplifies both at once (Figure~\ref{fig:dispersion}a): the mean
correlation rises as the names move together, and the cross-sectional
dispersion spikes as their magnitudes spread apart.

In the distance matrix the co-movement and the effective
dimension are two readings of one number. The participation ratio is
a tight, almost deterministic decreasing function of the mean
correlation, the same curve across all three periods, the two almost
perfectly anti-correlated near $-0.95$ (Figure~\ref{fig:dispersion}b):
as the cross section co-moves, its effective dimension collapses. In
the ranking chain the dispersion and the rank migration rise together
through the crash (Figure~\ref{fig:dispersion}c), the widening spread
of relative performance churning the order. The distance matrix and
the ranking are the two halves of the same daily cross section, the
systematic and the idiosyncratic, each amplified by a crisis.

\section{Discussion}
\label{sec:discussion}

The eigenvalue collapse at a crisis is a known fact, on its own a
scalar statement that correlations rise and the effective dimension
falls. The trajectory view refines it in three ways a single
correlation matrix cannot. The lookback comparison shows that the
standard two-year window is a low-pass filter that delays the onset
and hides precursors such as the mid-2007 tremor, while a short window
recovers them at the price of sampling noise that a market-cap cap
keeps under control. The market-factor removal shows the crisis is not
only a market-beta event, and that the sector geometry has its own
clock, most clearly in the 2015--2024 Covid period, where the sector
rotation runs a year or more past the crash. The rotation diagnostics
separate what the eigenvalue observables conflate, a growth of the
market eigenvalue at fixed direction from a genuine rotation of the
sector eigenbasis, with the random-subspace null certifying the
rotation as coherent.

The persistence of utilities as the most concentrated residual
cluster is worth a comment. After the market is removed, the
strongest remaining common factor is not one of the cyclical
sectors but the rate-sensitive utilities, which move together on
the level and slope of interest rates rather than on the
business cycle. The crisis-specific clusters, technology in 2001,
the energy supercycle in 2003--2012, and energy again in 2020,
appear as concentrations on top of this standing utilities block.

The forecasting test and the name-level attribution tell one
story from two sides. The fragility signals forecast only the
endogenously-building 2008 crisis, and the movers show why: in 2008
the energy cluster reorganizes gradually, giving the signal time to
rise, whereas in 2020 the rate-sensitive block breaks in a single
synchronized step at onset, with nothing to read in advance. The
synchronization of the 2020 movers is the name-level reason the ROC
finds no lead.

The distance matrix and the two ranking chains give a coherent
account from opposite directions. The distance matrix is dominated by
the market factor and reads a crisis as the collapse of the correlation
geometry onto that mode, while the ranking chains remove the market by
construction and read the same crisis as a churn of the relative
order, one chain for performance and one for risk. They agree on the
sector structure, the one signal each can see: the sector eigenvector
of the market-removed distance matrix and the within-sector
co-migration of the return chain both strengthen from 2001 to 2020.
On the crash itself they are complementary. A rise in average
correlation and a burst of rank migration are the co-movement and the
dispersion of the same violent move, the first visible only with the
market kept in the matrix, the second only after the ranking removes
it.

The two ranking chains are a matrix-form risk-return analysis in their
own right, sitting at opposite ends of the persistence scale, the
return chain forgetting an ordering in about seven days and the
volatility chain holding one for thirty to forty. Their entropy
production splits the same way, and both stay small, so the market is
at most weakly non-equilibrium in its relative order. The return chain
stays near the reversible bound, close to time-reversible, while the
volatility chain carries a weak, episodic arrow that flares at market
stress, reaching $z\approx 8$ at the 2002 dot-com bottom and
$z\approx 4$ at the 2008 crisis, placing the market's arrow of time in
the volatility as the Zumbach effect predicts. Their transfer-entropy
leadership splits too, the rate-driven utilities and other defensives
leading the return channel through the 2008 and Covid crises in a
ranking-space flight to quality, while the financial sector leads the
volatility channel, its risk propagating first. Those same utilities
anchor the market-removed residual as its most coherent cluster, so
coherence and the performance lead name one sector while the risk lead
names another.

\begin{figure}[htbp]
\centering
\includegraphics[width=0.98\textwidth]{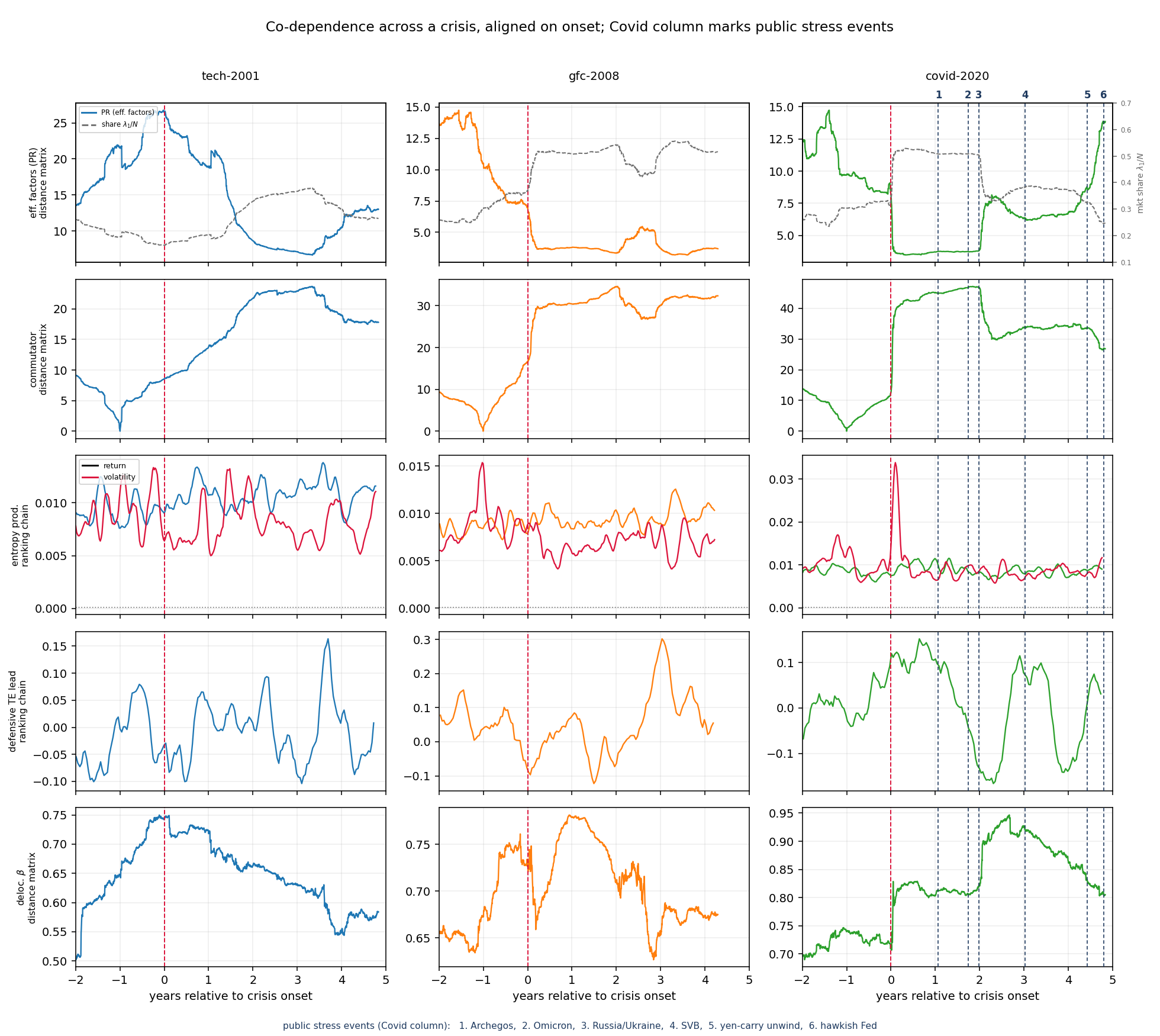}
\caption{Co-dependence of the matrix views across a crisis, aligned
on onset (crimson dashed at year zero), one column per period and out
to five years past onset. Rows, top to bottom: the effective factor
count (the participation ratio, PR) of the distance matrix, the
commutator norm, the entropy production of the ranking chain, the
defensive-sector transfer-entropy leadership, and the delocalised
exponent $\beta$ of the distance matrix. The first row superimposes
the market-factor share $\lambda_1/N$ (grey dashed, right axis) on the
PR, the same concentration read as a rising share and a falling factor
count. The entropy-production row draws the return chain in the period
colour and the volatility chain in red, its excess over the dotted
reversible null the time-irreversibility. At the 2008 and 2020 onsets
the distance-matrix signals fire at the crash and the ranking-chain
signals follow over the next months. In the Covid column the numbered
dashed lines mark public stress events (1 Archegos, 2 Omicron, 3
Russia/Ukraine, 4 SVB, 5 yen-carry unwind, 6 hawkish Fed).}
\label{fig:joint}
\end{figure}

These signals are coupled, but they are ordered in time rather than
simultaneous. Figure~\ref{fig:joint} aligns five of them on the
crisis onset for each period: the effective factor count (measured by
the participation ratio, PR), the commutator norm, and the delocalised
exponent $\beta$ from the distance matrix, and the entropy production
and the defensive transfer-entropy leadership from the ranking chain.
For the 2008 and 2020 crises the three distance-matrix signals fire at
the onset itself, the factor count collapsing, the commutator
stepping up, and $\beta$ steepening as the crash enters the window. The two ranking-chain
signals respond more slowly, over the following months, as the raw
entropy production lifts and the defensive sectors take over the
transfer-entropy leadership. The lag has a clear origin: the spectral
observables read the instantaneous correlation state, while the
ranking-chain quantities read a time-directional property that needs
both the fast crash and the slow recovery before it registers. The
five are still co-dependent readings of one reorganization of the
cross section, a loss of dimension, an eigenbasis rotation, a
steepening slope, a rise in irreversibility, and a shift of directed
influence toward the defensives. The 2001 bust, the dispersed regime,
is the exception in the distance-matrix rows, where the factor count
rises rather than falls, while the ranking rows still register the
event. The coupling across the two matrices, one built from
correlations and one from ranks, one keeping the market and one
removing it, is the strongest statement that they are watching the
same dynamics.

The diagnostics differ in what they are sensitive to, and a calendar
of documented market-stress events brings the difference out. A
sustained crash lights up the whole battery, dimension collapse and
eigenbasis rotation in the distance matrix, an entropy-production lift
and a leadership shift in the ranking chains. Brief single-session
shocks are another matter, and the arrow of time is the clearest case.
It fires on sustained, directionally asymmetric reshuffling, the 2002
and 2008 drawdowns and the multi-day forced deleveraging of the 2021
Archegos liquidation, and stays flat through one-day macro shocks such
as the Omicron scare, the SVB failure, or a hawkish-Fed repricing, even
when those move the VIX sharply. The selectivity is a feature rather
than a limitation. The entropy production measures directional
irreversibility, so it separates the events that leave a persistent,
time-asymmetric imprint on the cross section from those that spike
volatility and mean-revert within the window.

The spectral reading places the market next to the learning systems
the distance-matrix observable is borrowed from, and the contrast is
the sharpest statement of what the market does not do. In some
machine-learning experiments, grokking and sparse parity, the distance
matrix undergoes a geometric phase transition as the network learns:
its delocalised exponent rises through one, quasi-multiplets appear,
and a shoulder forms at $K\approx\sqrt N$ as the representation settles
onto a low-dimensional sub-manifold. The market's distance matrix does none of this. Its
exponent stays near $0.7$ across all three crises with no
quasi-multiplets or shoulder, and the crisis trajectories of this
paper are direct evidence that the geometry never settles. The market
carries the signature of an un-relaxed, high-dimensional
representation cloud, the state a learning system occupies before
training: it does not learn its correlation structure into a
low-dimensional geometry, and its distance matrix is a non-equilibrium
steady state through calm periods that a crisis drives further from
equilibrium rather than toward a new one. This turns the absence of
BBS structure from a negative finding into a positive one, and sets
the terms for the trajectory diagnostics that follow: the market is
watched not for signs of relaxation, which never come, but for how far
each crisis pushes an object with no equilibrium to return to.

Several limitations are worth stating. The role of the delocalised
exponent $\beta$ and the caveats on the strict BBS reading were set
out in Section~\ref{sec:observables} and are not repeated here. The
sector map is SIC-derived rather than GICS, so a
handful of names sit in a coarser bin than a practitioner
classification would assign, though the crisis-time leadership finding
rests on utilities, a clean SIC bin, and is not sensitive to the
ambiguous cases. The early-warning areas under the ROC are in-sample
and per-period, a descriptive comparison rather than an out-of-sample
forecast, so their regime-dependence is qualitative. The market-cap
cap used for the short lookbacks changes the universe between
lookbacks, so the 126 and 504-day curves are not on identical name
sets, though the crisis-relative changes are insensitive to the cap.
Finally, the anchor for the rotation diagnostics is a fixed pre-crisis
date, so the reported drifts and commutators are relative to a calm
reference and are not meant as an absolute scale.

\section{Summary and outlook}
\label{sec:summary}

We read the S\&P 500 cross section through three fixed-size matrix
trajectories. The arccos distance matrix reproduces the known
spectral collapse at crises and adds three things beyond it: a
sharper, precursor-resolving onset at short lookback, a sub-market
sector reorganization with its own timing under market removal, and
a coherent rotation of the sector eigenbasis whose composition names
each crisis by industry. The same trajectory-level diagnostics that
separate a coherent rotation from random tumbling in a physical
relaxation problem \cite{halperin2026FDM} carry over unchanged. The
other two matrices order the names each day by return and by
volatility, two market-neutral Markov chains that read the relative
dynamics the market factor hides. They are a matrix-form risk-return
analysis. The return chain mixes in about a week and the volatility
chain in a month or more, and their transfer-entropy leadership,
resolved down to individual bellwethers, splits between the defensive
utilities that lead the performance channel through the 2008 and
Covid crises and the financial sector that leads the risk channel.
Their entropy production splits the same way, both small in absolute
terms. The return chain stays near the reversible bound at all times,
so relative performance is close to time-reversible, while a weak,
episodic arrow of time surfaces in the volatility chain, the
ranking-space form of volatility clustering and the Zumbach effect.
Resolved quarterly across thirty years, the volatility arrow is near
zero in calm markets and flares at market-stress episodes, strongest
at the 2002 dot-com bottom and the 2008 crisis. This episodic,
stress-driven irreversibility, small but genuine, is a reading the
discrete Markov chains make accessible where a continuous,
high-dimensional estimate would struggle.

Read through the same spectral lens as the machine-learning systems
the distance-matrix observable comes from, the market never relaxes.
Its delocalised exponent stays near $0.7$ across all three crises with
no quasi-multiplets and no shoulder, the signature of an un-relaxed,
high-dimensional representation cloud rather than a geometry settled
onto a low-dimensional sub-manifold. Where a learning system passes
through a geometric phase transition and its distance matrix organises,
the market does not, its distance matrix a permanently non-equilibrium
object that a crisis drives further from equilibrium rather than toward
a new steady state.

The three views are consistent and complementary, the systematic and
the idiosyncratic halves of one cross section. The spectral collapse
fires at the crash and the ranking responses follow over the next
months, and the rate-driven utilities block that anchors the
market-removed residual is the same one that leads the performance
channel of the ranking transfer entropy.

The results give a consistent picture of what these diagnostics can
identify and what they can predict. Crises are identifiable across the
whole battery: the distance matrix loses dimension and rotates its
eigenbasis, while the ranking chains lift their entropy production and
shift their leadership. The individual market-stress events of the
Covid decade are identifiable too, and which signal moves classifies
the event, the arrow of time responding to a directional deleveraging
like Archegos and the spectral concentration to a correlation shock
like the Silicon Valley Bank failure, while brief symmetric shocks
such as the Omicron scare leave little trace. Predictability is the
harder half. Only the endogenously-building crisis, the 2008 buildup,
is forecast in advance by the fragility signals. The exogenous 2020
crash and the single-session stress events are detected as they
happen, not before. The diagnostics are thus a strong contemporaneous
read of the cross section and only a partial early warning, informative
ahead of time when a crisis assembles slowly enough to leave a
fragility trail.

The approach suggests several natural extensions. The first is to
add covariates to the ranking-chain dynamics. We treated the two
chains as memoryless as a first approximation, and the systematic
sector and firm-specific factors, together with the higher-order
memory of the transitions, can be built in as exogenous drivers of
the transition matrix. A state-dependent chain of this kind would add
realism. Coupled to the persistence and leadership structure
documented here, it points toward dynamic portfolio management, where
a rank-based or transfer-entropy-informed strategy that buys the
followers when the defensives lead connects these observables to
stochastic portfolio theory \cite{fernholz2002,banner2005}.

More broadly, the matrix-based formalism itself applies to portfolio
construction and optimization. The effective dimension and
correlation geometry of the distance matrix inform covariance
estimation and risk budgeting, and the persistence, mixing time, and
directed leadership of the ranking chains inform regime-conditioned
allocation and rank-based tilts. The natural target is a dynamic
portfolio-optimization problem whose objective adapts to the measured
fragility and leadership state, with the fixed-size matrix observables
serving as the state variables that carry the market history.

Two further refinements concern the early-warning reading of
Section~\ref{subsec:earlywarning}. The forecastable regime can be
separated cleanly by conditioning on whether the correlation structure
is already tightening, rather than pooling all crises into one ROC.
The name-level attribution of Section~\ref{subsec:names} can be
recomputed at a shorter lookback, trading a smaller large-cap universe
for a tighter onset, and folded into a real-time count of how many
names have already moved, a coincidence indicator alongside the
fragility levels.


\begin{thebibliography}{99}
\small
\setlength{\itemsep}{0pt}
\setlength{\parskip}{0pt}
\setlength{\parsep}{0pt}

\bibitem{marchenko1967}
V.~A.~Mar\v{c}enko and L.~A.~Pastur.
\emph{Distribution of Eigenvalues for Some Sets of Random
Matrices.}
\textit{Mathematics of the USSR-Sbornik} \textbf{1}, 457--483
(1967).

\bibitem{laloux1999}
L.~Laloux, P.~Cizeau, J.-P.~Bouchaud, and M.~Potters.
\emph{Noise Dressing of Financial Correlation Matrices.}
\textit{Physical Review Letters} \textbf{83}, 1467--1470 (1999).

\bibitem{plerou2002}
V.~Plerou, P.~Gopikrishnan, B.~Rosenow, L.~A.~N.~Amaral,
T.~Guhr, and H.~E.~Stanley.
\emph{Random Matrix Approach to Cross Correlations in Financial
Data.}
\textit{Physical Review E} \textbf{65}, 066126 (2002).

\bibitem{potters2020}
M.~Potters and J.-P.~Bouchaud.
\emph{A First Course in Random Matrix Theory: for Physicists,
Engineers and Data Scientists.}
Cambridge University Press, Cambridge (2020).

\bibitem{onnela2003}
J.-P.~Onnela, A.~Chakraborti, K.~Kaski, J.~Kert\'esz, and
A.~Kanto.
\emph{Dynamics of Market Correlations: Taxonomy and Portfolio
Analysis.}
\textit{Physical Review E} \textbf{68}, 056110 (2003).

\bibitem{kritzman2011}
M.~Kritzman, Y.~Li, S.~Page, and R.~Rigobon.
\emph{Principal Components as a Measure of Systemic Risk.}
\textit{Journal of Portfolio Management} \textbf{37}(4),
112--126 (2011).

\bibitem{halperin2026omd}
I.~Halperin.
\emph{Learning as Observable Matrix Dynamics: Diffusive
Relaxations versus Phase Transitions.}
arXiv:2606.29679, 2026.

\bibitem{halperin2026IBBS}
I.~Halperin.
\emph{I-BBS: Coordinate-Free Inference of Latent Sub-Manifolds
Using Random Distance Matrix Theory.}
arXiv:2606.29675, 2026.

\bibitem{bogomolny2003}
E.~Bogomolny, O.~Bohigas, and C.~Schmit.
\emph{Spectral Properties of Distance Matrices.}
\textit{Journal of Physics A: Mathematical and General}
\textbf{36}, 3595--3616 (2003). arXiv:nlin/0301044.

\bibitem{bogomolny2007}
E.~Bogomolny, O.~Bohigas, and C.~Schmit.
\emph{Distance Matrices and Isometric Embeddings.}
arXiv:0710.2063, 2007.

\bibitem{mezard1999}
M.~M\'ezard, G.~Parisi, and A.~Zee.
\emph{Spectra of Euclidean Random Matrices.}
\textit{Nuclear Physics B} \textbf{559}, 689--701 (1999).

\bibitem{goetschy2013}
A.~Goetschy and S.~E.~Skipetrov.
\emph{Euclidean Random Matrices and Their Applications in
Physics.}
arXiv:1303.2880, 2013.

\bibitem{halperin2026FDM}
I.~Halperin.
\emph{Frustrated Dynamics of Distance Matrices.}
arXiv:2605.05376, 2026.

\bibitem{paninski2003}
L.~Paninski.
\emph{Estimation of Entropy and Mutual Information.}
\textit{Neural Computation} \textbf{15}, 1191--1253 (2003).

\bibitem{buth2025infomeasure}
C.~M.~B\"uth, K.~Acharya, and M.~Zanin.
\emph{infomeasure: A Comprehensive Python Package for Information
Theory Measures and Estimators.}
\textit{Scientific Reports} \textbf{15}, 14053 (2025).
arXiv:2505.14696.

\bibitem{wollstadt2019idtxl}
P.~Wollstadt, J.~T.~Lizier, R.~Vicente, C.~Finn, M.~Mart\'{\i}nez-Zarzuela,
P.~Mediano, L.~Novelli, and M.~Wibral.
\emph{IDTxl: The Information Dynamics Toolkit xl.}
\textit{Journal of Open Source Software} \textbf{4}(34), 1081 (2019).

\bibitem{jackwerth2018relative}
J.~C.~Jackwerth and A.~Slavutskaya.
\emph{Relative Alpha.}
Working paper, University of Konstanz, SSRN 2439145 (2018).

\bibitem{forsberg2021peer}
D.~Forsberg, D.~R.~Gallagher, and G.~J.~Warren.
\emph{Identifying Hedge Fund Skill by Using Peer Cohorts.}
\textit{Financial Analysts Journal} \textbf{77}(2), 97--123 (2021).

\bibitem{fernholz2002}
E.~R.~Fernholz.
\emph{Stochastic Portfolio Theory.}
Springer, New York (2002).

\bibitem{banner2005}
A.~D.~Banner, R.~Fernholz, and I.~Karatzas.
\emph{Atlas Models of Equity Markets.}
\textit{Annals of Applied Probability} \textbf{15}(4),
2296--2330 (2005).

\bibitem{seifert2005}
U.~Seifert.
\emph{Entropy Production along a Stochastic Trajectory and an
Integral Fluctuation Theorem.}
\textit{Physical Review Letters} \textbf{95}, 040602 (2005).

\bibitem{seifert2012}
U.~Seifert.
\emph{Stochastic Thermodynamics, Fluctuation Theorems and
Molecular Machines.}
\textit{Reports on Progress in Physics} \textbf{75}, 126001
(2012).

\bibitem{georgiev2025}
G.~Georgiev.
\emph{Stochastic Thermodynamics of Financial Markets I: Entropy
Production and the Second Law.}
SSRN preprint 5454994, 2025.

\bibitem{halperin2026entropy}
I.~Halperin.
\emph{Information Bottleneck, Semantic Faithfulness and Entropy
Production Measures for Analysis of Large Language Models as
Information Engines.} 2026.

\bibitem{parrondo2009}
J.~M.~R.~Parrondo, C.~Van den Broeck, and R.~Kawai.
\emph{Entropy Production and the Arrow of Time.}
\textit{New Journal of Physics} \textbf{11}, 073008 (2009).

\bibitem{zumbach2009}
G.~Zumbach.
\emph{Time Reversal Invariance in Finance.}
\textit{Quantitative Finance} \textbf{9}(5), 505--515 (2009).

\bibitem{schreiber2000}
T.~Schreiber.
\emph{Measuring Information Transfer.}
\textit{Physical Review Letters} \textbf{85}, 461--464 (2000).

\bibitem{kraskov2004}
A.~Kraskov, H.~St\"ogbauer, and P.~Grassberger.
\emph{Estimating Mutual Information.}
\textit{Physical Review E} \textbf{69}, 066138 (2004).

\end{thebibliography}
\end{document}